\documentclass[twocolumn,aps]{revtex4-1}

\usepackage{amsmath}
\usepackage{soul}

\usepackage{amssymb}
\usepackage{graphicx}
\usepackage{color}
\usepackage{bm}

\newcommand{\be}{\begin{equation}}
\newcommand{\ee}{\end{equation}}

\newcommand{\benn}{\begin{equation*}}
\newcommand{\eenn}{\end{equation*}}

\newcommand{\bea}{\begin{eqnarray}}
\newcommand{\eea}{\end{eqnarray}}

\newcommand{\ve}{\varepsilon}
\newcommand{\mA}{\mathcal{A}}

\newcommand{\mC}{\mathcal{C}}

\newcommand{\mO}{\mathcal{O}}
\newcommand{\mH}{\mathcal{H}}
\newcommand{\mI}{\mathcal{I}}

\newcommand{\mZ}{\mathcal{Z}}

\newcommand{\mycolor}[1]{{\color{black} #1}}
\newcommand{\mycol}[1]{{\color{black} #1}}

\newcommand{\gcol}[1]{{\color{black} #1}}
\newcommand{\ggcol}[1]{{\color{black} #1}}

\usepackage{bm}
\usepackage{tikz}
\usetikzlibrary{calc}
\usetikzlibrary{shapes}
\usetikzlibrary{calc,decorations.markings,decorations.pathmorphing}
\tikzset{cross/.style={cross out, draw=black, minimum size=2*(#1-\pgflinewidth), inner sep=0pt, outer sep=0pt},cross/.default={1pt}}

\begin{document}

\title{ \ggcol{Localization transition in the Discrete Non-Linear
    Schr\"odinger Equation:\\ ensembles
    inequivalence and negative temperatures.}}

\author{Giacomo Gradenigo} \affiliation{Gran Sasso Science Institute,
  Via F. Crispi 7, 67100 L'Aquila, Italy\\ NANOTEC-CNR, Roma, Piazzale A. Moro 2, I-00185
  Roma, Italy}

\author{Stefano Iubini}
\affiliation{Consiglio Nazionale delle Ricerche, Istituto dei Sistemi Complessi, via Madonna del Piano 10, I-50019 Sesto Fiorentino, Italy}
\affiliation{Istituto Nazionale di Fisica Nucleare, Sezione di Firenze, via G. Sansone 1 I-50019, Sesto Fiorentino, Italy}

\author{Roberto Livi}
\affiliation{Dipartimento di Fisica e Astronomia and CSDC, Universit\`a degli Studi di Firenze \&\\ Istituto Nazionale di Fisica Nucleare,
  Sezione di Firenze, via G. Sansone 1 I-50019, Sesto Fiorentino, Italy}
 \affiliation{Consiglio Nazionale delle Ricerche, Istituto dei Sistemi Complessi, via Madonna del Piano 10, I-50019 Sesto Fiorentino, Italy }

\author{Satya N. Majumdar} \affiliation{LPTMS, CNRS, Universit\'e Paris-Sud,
  Universit\'e Paris-Saclay, 91405 Orsay, France}

\begin{abstract}

  We present a detailed account of a first-order localization
  transition in the Discrete Nonlinear Schr\"odinger Equation, where
  the localized phase is associated to the high energy region in
  parameter space.  We show that, due to ensemble inequivalence, this
  phase is thermodynamically stable only in the microcanonical
  ensemble. In particular, we obtain an explicit expression of the
  microcanonical entropy close to the transition line, located at
  infinite temperature. This task is accomplished making use of
  large-deviation techniques, that allow us to compute, in the limit
  of large system size, also the subleading corrections to the
  microcanonical entropy.  These subleading terms are crucial
  ingredients to account for the first-order mechanism of the
  transition, to compute its order parameter and to predict the
  existence of negative temperatures in the localized phase. All of
  these features can be viewed as signatures of a thermodynamic phase
  where \ggcol{the translational symmetry is broken spontaneously due
    to} a condensation mechanism yielding energy fluctuations far away
  from equipartition: actually they prefer to participate in the
  formation of nonlinear localized excitations (breathers), typically
  containing a macroscopic fraction of the total energy.

 \end{abstract}

\maketitle


\section{Introduction}
\label{sec:0}

\begin{figure}
	  \includegraphics[width=0.9\columnwidth]{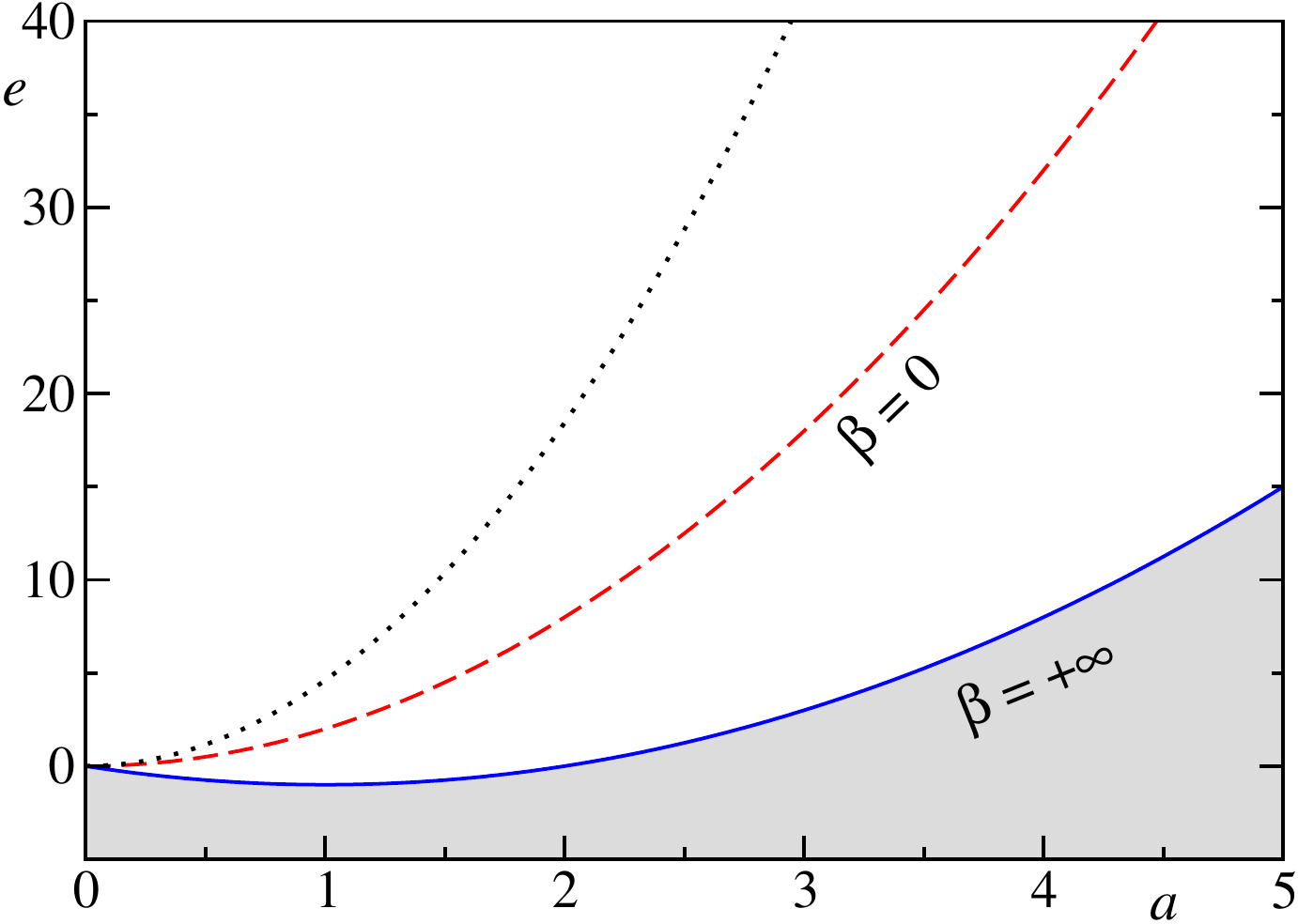}
  \caption{Equilibrium phase diagram in the plane $(a=A/N, e =E/N)$
    for the DNLSE as obtained in~\cite{RCKG00}.  The ground state
    $(\beta=\infty)$ is identified by the curve $e=a^2-2a$: the grey
    area below this curve is inaccessible.  The infinite temperature
    line, $\beta=0$, which corresponds to the parabola $e = 2a^2$, is
    the border of the region where ensembles are equivalent, and
    represents also the transition line for the localization phase
    transition in the $N\rightarrow\infty$ limit. For finite $N$, the
    localization transition occurs above the $\beta=0$ line: black
    dotted curve represents the transition line for $N=100$.}
\label{fig1}
\end{figure}

Beyond its phenomenological interest for many physical applications,
such as Bose-Einstein condensates in optical lattices \cite{TS,FLOP}
or light propagating in arrays of optical waveguides \cite{ESMBA}, in
the last decades the Discrete Nonlinear Schr\"odinger Equation (DNLSE)
has revealed an extremely fruitful testbed for many basic aspects
concerning statistics and dynamics in Hamiltonian models, equipped
with additional conserved quantities, other than energy (e.g., see
\cite{JR} and \cite{LFO}).  In particular, since the pioneering paper
by Rasmussen et al.~\cite{RCKG00}, the existence of a high-energy
phase characterized by the condensation of energy in the form of
breathers, i.e. localized nonlinear excitations, has attracted the
attention of many scholars. The phase diagram shown in Fig.~\ref{fig1}
summarizes the scenario described in \cite{RCKG00}. From a dynamical
point of view, extended numerical investigations have pointed out that
in such a breather-rich high energy phase the Hamiltonian evolution of an isolated chain
exhibits long-living multi-breather states, that last over
astronomical times~\cite{FLOP,ICOPP}.  Moreover, time averages of
suitable non-local quantities entering the definition of temperature
$T$, as, e.g., the inverse of the derivative of the microcanonical
entropy $S$ with respect to the energy $E$ (i.e., $\beta =1/T =
\frac{\partial S} {\partial E}$)~\cite{Franzosi}, predict that $T$ is
negative in this high-energy phase~\cite{IFLOP}.  Relying upon
thermodynamic considerations Rumpf and Newell~\cite{RN} and later
Rumpf \cite{R1,R2,R3,R4} argued that these dynamical states should
eventually collapse onto an ``equilibrium state'', characterized by an
extensive background at infinite temperature with a superimposed
localized breather, containing the excess of energy initially stored
in the system. This argument seems plausible in the light of a
grand-canonical description, where entropy has to be maximal at
thermodynamic equilibrium, in the presence of fluctuations due to heat
exchanges with a thermal reservoir. Accordingly, in this perspective
long-living multi-breather states were interpreted as metastable ones,
thus yielding the conclusion that negative temperature states are not
genuine equilibrium ones.  Due to the practical difficulty of
observing the eventual collapse to the equilibrium state predicted by
Rumpf even in lattices made of a few tens of sites, some authors have
substituted the Hamiltonian dynamics with a stochastic evolution rule
that conserves energy and particle densities.  In this framework the
breather condensation phenomenon onto a single giant breather emerges
as a coarsening process ruled by predictable scaling properties
\cite{IPP14,IPP17}.

Despite the fact that many of the dynamical aspects of the deterministic and
stochastic evolutions of the DNLSE have been satisfactorily described
and understood, the thermodynamic interpretation of this model
still remains unclear as it seems to depend on the choice of the
statistical ensemble.  In fact, already in \cite{RCKG00} the
authors point out that an approach based on the canonical ensemble
yields a negative temperature in the high-energy phase, which
contradicts the very existence of a Gibbsian measure. Moreover, they
guess that a consistent definition of negative temperatures compatible
with a grand-canonical
representation could be obtained  at the price of transforming the original short-range
Hamiltonian model into a long-range one.  Also Rumpf makes use of a
grand-canonical ensemble to tackle this question \cite{R3,R4} and
reaches the opposite conclusion that negative temperature states are
not compatible with thermodynamic equilibrium conditions.  More
recently, the statistical mechanics of the disordered DNLSE
Hamiltonian has been analyzed making use of the grand-canonical
formalism \cite{BM18}: the authors conclude that for weak disorder the
phase diagram looks like the one of the non-disordered model, while
correctly pointing out that their results apply to the microcanonical
case, whenever the equivalence between ensembles can be established.
In a more recent paper the thermodynamics of the DNLSE Hamiltonian and
of its quantum counterpart, the Bose-Hubbard model, has been analyzed
in the canonical ensemble~\cite{CEF}: the authors even claim that
``... the Gibbs canonical ensemble is, conceptually, the most
convenient one to study this problem'' and conclude that the
high-energy phase is characterized by the presence of non-Gibbs
states, that cannot be converted into standard Gibbs states by
introducing negative temperatures. The main consideration emerging
from the overall scenario is that, while in the low-energy phase the
thermodynamics of the DNLSE model exhibits standard properties that
are consistent with any statistical ensemble representation, in the
high-energy phase the very equivalence between statistical ensembles
is at least questionable and it cannot be ruled out as a matter of
taste.\\

In this paper we present a clear scenario about the statistical
mechanics of the DNLSE by performing explicit analytic
calculations within the microcanonical ensemble in the high energy limit
and, in particular, close to the $\beta = 0$ line (see
Fig.~\ref{fig1}).
We show explicitly that
the non-equivalence between statistical ensembles naturally emerges as
a consequence of the non-analytic structure of the
microcanonical partition function in the high-energy phase.
 Moreover, we find  that
a first-order phase transition between a
low-energy extensive  phase and a high-energy localized phase  occurs
 close to the $\beta=0$ line and we include  explicit
predictions about the scaling of the partition function and finite
size corrections.
We analyze the transition by computing the
participation ratio of the energy per lattice site as  an appropriate
indicator.
Finally, we obtain the explicit form of the
microcanonical entropy and we show
that temperatures are  negative for any finite system size in the region of ensemble inequivalence.

The paper is organized as follows. In Sec.~\ref{sec:1} we briefly
comment the state of the art of the DNLSE thermodynamics, mainly by
the results of Ref.~\cite{RCKG00}. We introduce also the main object
of this study: the microcanonical partition function of the DNLSE,
discussing the only approximation done through the whole paper,
i.e. the neglecting of the hopping terms close to the region at
infinite temperature.  In Sec.~\ref{sec:2} we show how to cast the
calculation of the microcanonical partition function as a
large-deviation calculation, i.e. as the calculation of the
probability distribution of independent identically distributed random
variables. In particular we show how to account for the
\mycol{non-analytic} contribution to the microcanonical partition
function, coming from a cut in the negative real semiaxis in the
complex $\beta$ domain of the canonical partition function. Then in
Sec.~\ref{sec:3} we explain how the non-analytic part of the partition
function gives rise to a discontinuous contribution to the first
derivative of the microcanonical entropy and we also compute the
participation ratio of the DNLSE, which is the natural order parameter
of this localization first-order transition. Moreover, we explain how
negative temperatures arise naturally in this context and finally, in
Sec.~\ref{sec:4}, we turn to conclusions.  The technical details of
the calculations are contained in the appendices.

\section{The Discrete Non-Linear Schr\"odinger equation: generalities}
\label{sec:1}
The DNLSE is a one-dimensional model of a scalar
complex field $z_j$ on a lattice made of $N$ sites \mycol{with periodic
boundary conditions} , whose Hamiltonian reads
\be
\mH = \sum_{j=1}^N (z_j^*z_{j+1}+z_j z^*_{j+1}) + \sum_{j=1}^N |z_j|^4 \,.
\label{eq:Hamiltonian}
\ee
The corresponding Hamiltonian dynamics is expressed by the equations of motion
\be
i \dot{z_j} = - \frac{\partial \mH}{\partial z_j^*} =
-(z_{j+1}+z_{j-1})- 2 |z_j|^2 z_j \,.
\label{eq:eq-motion}
\ee

It is crucial to note that these equations of motion
conserve not only the total energy $\mH$, but,
due to the `quantum' origin of the problem, also the squared norm of the
``wave-function'', that is
\be
\mA = \sum_{j=1}^N |z_j|^2 \,.
\label{norm.1}
\ee
This quantity $\mA$ can also be interpreted as the total number of particles
present in the system. We attribute to $\mH$ and $\mA$ the real
values $E$ and $A$, respectively.  

The phase diagram of this model can be 
easily characterized by representing the complex field in its polar form
$z_j = e^{i \phi_j}\rho_j$, so that (\ref{eq:Hamiltonian}) can be rewritten as
\be
\mathcal{H} = 2 \sum_{j=1}^N \rho_j\rho_{j+1} \cos(\phi_{j+1}-\phi_j)+ \sum_{j=1}^N \rho_j^4.
\label{eq:H-angles}
\ee It is straightforward to realize that the ground state of this
Hamiltonian is obtained when $\cos(\phi_{j+1}-\phi_j)=-1$ for all $j$,
and $\rho_j=\rho$, independently of the site index $j$.  In terms of
the energy and mass densities $e = E/N$ and $a = A/N$, the ground
state is identified by the condition $e= -2\, a+ a^2$, \gcol{since one
  has $\rho^2=a$}. This line is shown in Fig.~\ref{fig1} and in a
thermodynamic interpretation it corresponds to the condition $\beta =
1/T=+\infty$.  Note that the region below this line is not physically
accessible. The other important line characterizing the phase diagram
corresponds to infinite temperature, i.e. $\beta =0$.  One can argue
that in this case the angular variables $\lbrace
\phi_j\rbrace_{j=1,\ldots,N}$ have to behave as independent
identically distributed (i.i.d) random variables, so that the first
addendum in the r.h.s. of Eq.~(\ref{eq:H-angles}) is proportional to
$\sqrt{N}$, while the second one is proportional to
$N$. \gcol{Therefore the contribution of the bilinear terms to the
  total energy is subleading and we will neglect it in our calculation
  of the partition function, done in the spirit of a large-$N$
  estimate. At high temperatures the subleading nature of the bilinear
  terms, which should give a term proportional to $a$, is confirmed by
  the exact analytical result of~\cite{RCKG00}.  In~\cite{RCKG00} such
  bilinear terms are taken into account and, despite this, the
  $\beta=0$ line is still identified by the condition $e=2\, a^2$ (see
  Fig.~\ref{fig1}): that is, there is no term linear in $a$.}

Due to the presence of two conserved quantities, the deterministic
dynamics of this model exhibits quite peculiar features.  In the
region sandwitched between the zero temperature line and the infinite
temperature one, any initial condition relaxes to a uniform
equilibrium state displaying equipartition.  Conversely, above the
($\beta=0$)-line the dynamics on any finite lattice is characterized
by the birth and death of long-living localized nonlinear excitations,
called breathers (e.g. see \cite{TS,FLOP}).  In a stochastic version
of the dynamics~\cite{IPP14,IPP17} still conserving both $E$ and $A$,
the evolution in the region above the ($\beta=0$)-line has been found
to lead to a coarsening process eventually yielding a state where a
giant breather confined in a finite lattice region collects a finite
fraction of the total energy and norm (number of particles). In
particular, the condensate \mycolor{``mass''} has been found to
increase in time $t$ as $t^{1/3}$.  One cannot exclude that also the
deterministic dynamics in the localized phase might eventually evolve
to this single-breather state, even though it has never been observed
even in relatively small lattices.

Finding an interpretation of this scenario is related to the
understanding of the thermodynamics of the DNLSE model. Various
attempts have been made to study its equilibrium properties in the
presence of a localized wave-function ~\cite{RCKG00,BM18}.  The reason
why a study of the thermodynamics {\em deep} in the localized phase
has been (to some extent) so far elusive is that the ($\beta=0$)-line,
where condensation takes place, corresponds also to the breakdown of
the equivalence between statistical ensembles, as we are going to show
in Sec. \ref{sec:2-B}.  In this respect, it is worth quoting a series
of recent papers, focused on the regression dynamics of large
anomalous fluctuations, where it has been argued that isolated
anomalous fluctuations should be somehow related to ensemble
inequivalence~\cite{C15,C17,CMG19}. Nevertheless, \gcol{we need to
  stress that our present work demonstrates in a clear and transparent
  manner that the physics of the localized phase in the DNLSE can be
  described consistently {\it only in the microcanonical ensemble}.\\}

In fact, the main goal of this paper is the computation of  the microcanonical
entropy of the DNLSE Hamiltonian (\ref{eq:Hamiltonian}) for fixed
values of $E$ and $A$ close to the ($\beta=0$)-line, shown in
Fig.~\ref{fig1}. In general, the microcanonical entropy is defined as
\be
\mycol{S_N(A,E) = \log~\Omega_N^{O}(A,E)},
\ee
where the Boltzmann constant $k_B$ is set to unit  and $\Omega_N^O(A,E)$ is the microcanonical partition function:
\begin{widetext}
\be
\mycol{\Omega_N^{O}(A,E)} = \int \prod_{j=1}^N ~ d\mu(z_j) ~ \delta\left(A-\sum_{j=1}^N |z_j|^2\right)~ \delta\left(E-\left[\sum_{j=1}^N (z_j^*z_{j+1}+z_j z^*_{j+1})+\sum_{j=1}^N |z_j|^4\right] \right).
\label{eq:microcanonical-hop}
\ee
\end{widetext}
Here $ d\mu(z_j)$ is a shorthand notation for $d[\Re(z_j)] \, d[ \Im(z_j)]$.

Since the main interesting features of the DNLSE model occur in the
vicinity of the ($\beta=0$)-line, for large values of $N$, as discussed above,
the contribution to the total energy $E$ of the bi-linear
hopping term in Eq.~(\ref{eq:Hamiltonian}) can be neglected with
respect to the local nonlinear term. 
Accordingly, close to $\beta =0$ the microcanonical partition
function can be approximated as follows:
\begin{align}
  & \mycol{\Omega_N(A,E)} = \nonumber \\
  & \int \prod_{i=1}^N ~ d\mu(z_j) ~ \delta\left(A-\sum_{j=1}^N |z_j|^2\right)~ \delta\left(E-\sum_{j=1}^N |z_j|^4 \right).
\label{eq:microcanonical}
\end{align}
We want to remark that a similar microcanonical partition function with
two constraints was also studied~\cite{SEM14,SEM14b,SEM17} in the context of generalised
mass transport models on a lattice~\cite{MEZ05,EMZ06,EH2005,S08-leshouches}.

\section{The microcanonical partition function}
\label{sec:2}

In this Section we present the explicit computation of the
microcanonical partition function $\Omega_N(A,E)$ and, in particular,
its large deviation properties in the limit of large $N$.

\subsection{From statistical mechanics to large deviations}
\label{sec:2-A}

In order to compute $\mycol{\Omega_N(A,E)}$ it is convenient to
express the variables in their polar form $z_j=\rho_j e^{i\phi_j}$,
thus making straightforward the integration over the angular
coordinates $\phi_i$:
\begin{align}
  & \mycol{\Omega_N(A,E)} = \nonumber \\
  & (2\pi)^N \int_0^\infty \left[ \prod_{j=1}^N d\rho_j~\rho_j\right]~\delta\left(A - \sum_{j=1}^N
\rho_j^2\right)~\delta\left(E-\sum_{j=1}^N \rho_j^4\right).
\label{eq:microcanonical_2}
\end{align}
The effective strategy for proceeding in this analytic calculation
amounts to releasing the conservation constraint on $A$ \mycol{by
  means of a Laplace transform:}
\begin{align} & \tilde\Omega_N(\lambda,E) = \int_0^\infty dA~e^{-\lambda
  A}~\mycol{\Omega_N(A,E)} = \nonumber \\ & (2\pi)^N
\int_0^\infty~\left[\prod_{j=1}^N d\rho_j~\rho_j\right]~ e^{-
  \lambda\sum_{j=1}^N \rho_j^2}~~\delta\left( E -\sum_{j=1}^N \rho_j^4
\right).
\label{eq:partition-ld0}
\end{align}
From Eq.~(\ref{eq:partition-ld0}) the partition function at
fixed $A$ can be obtained by a simple inversion of the Laplace
transform:
\be
\Omega_N(A,E) = \frac{1}{2\pi i} \int_{\lambda_0-i\infty}^{\lambda_0+i\infty} d\lambda~e^{\lambda A}~\tilde\Omega_N(\lambda,E),
\label{eq:invL-A}
\ee
where integration goes over a vertical Bromwich contour in the complex
$\lambda$ plane, crossing the real $\lambda$ axis at a $\lambda_0$
which can be chosen to the right of all singularities of the
integrand. \gcol{But, for the moment, let us move forward to the
  calculation of $\tilde\Omega_N(\lambda,E)$.} By introducing the
change of variables $\rho_j^4 = \ve_j$ in
Eq.~(\ref{eq:partition-ld0}), it is useful to re-write
$\tilde\Omega_N(\lambda,E)$ as
\begin{align}
  & \tilde\Omega_N(\lambda,E) = \nonumber \\
  & \left(\frac{\pi}{2}\right)^N~\int_0^\infty~\prod_{j=1}^N
\frac{d\ve_j}{\sqrt{\ve_j}}~e^{- \lambda\sum_{j=1}^N \sqrt{\ve_j}}~\delta\left( E -\sum_{j=1}^N \ve_j \right).
\label{eq:canonical-1}
\end{align}
We can introduce the normalized probability distribution function (PDF)
\be
f_\lambda(\ve) = \Theta(\ve)~\frac{\lambda}{2}\frac{1}{\sqrt{\ve}}\exp\left( -\lambda\sqrt{\ve} \right),
\label{eq:fx}
\ee
where $\Theta(\ve)$ is the Heavyside distribution, and rewrite
\be
\tilde\Omega_N(\lambda,E) = \left(\frac{\pi}{\lambda}\right)^N~\mZ_N(\lambda,E)
\label{eq:canonical-2}
\ee
with
\be
\label{eq:semipart}
\mZ_N(\lambda,E) = \int_0^\infty~\prod_{j=1}^N~d\ve_j~f_\lambda(\ve_j)~\delta\left( E -\sum_{j=1}^N \ve_j \right),
\ee
In this form, $\mZ_N(\lambda,E)$ can be interpreted as the probability
distribution of \mycol{the sum of} $N$ i.i.d. random variables
$\ve_j$, each drawn from the PDF $f_\lambda(\ve_j)$.  
Note that $\lambda$ appears
only as a
parameter in the PDF $f_\lambda(\ve_j)$ in Eq. (\ref{eq:fx}).

It is well known
from the theory of large deviations~\cite{MEZ05,EMZ06,S08-leshouches}
that the global constraint on the sum of the variables yields a
condensation phenomenon when the individual probability distribution
is {\sl fat-tailed}, i.e., when it fulfills the following bounds for
large $\ve$
\be
\exp(-\ve)< \mycolor{f_\lambda(\ve)} < \frac{1}{\ve^2} \,,
\label{efp}
\ee
which is exactly the case of Eq.~(\ref{eq:fx}).  More precisely, when
the sum of these i.i.d. random variables (in our case the total energy
$E$) overtakes a threshold value (that in our case we denote
$E_{\text{th}}$) one observes a crossover from a ``fluid phase'',
where the sum is democratically shared among all variables
(i.e. energy is equally distributed at each lattice position for
$E<E_{\text{th}}$) to a ``condensed phase'', where the behaviour of
$\mZ_N(\lambda,E)$ in (\ref{eq:semipart}) is dominated by the
probability distribution $f_\lambda(\ve_i)$ of a single variable
(i.e., a macroscopic fraction of the total energy localizes at some
lattice site for $E>E_{\text{th}}$). Thus this `condensed' phase
precisely corresponds to the `localised' phase in the DNLSE model.
The value of $E_{\text{th}}$ is
known to be equal to the average total energy, i.e.,
\be
\label{thres}
E_{\text{th}} = N\, \langle \ve \rangle \,,
\ee
where the average $\langle \,\,\,\rangle$ is over the probability
distribution $f_\lambda(\ve)$~\cite{S08-leshouches}.  Making use of Eq.~(\ref{eq:fx}) one 
obtains
\bea
\langle \ve \rangle &=& \frac{2}{\lambda^2} \nonumber \\
\langle \ve^2 \rangle &=& \frac{24}{\lambda^4} \nonumber \\
\sigma^2 &=& \langle \ve^2 \rangle - \langle \ve \rangle^2 = \frac{20}{\lambda^4}
\label{eq:moments-1-2}
\eea
These results will be used also in the following sections. 
We anticipate here that the explicit computation of the microcanocal
partition function yields $\lambda=1/a$ (see Eq.(\ref{spc})~) so that we can write 
$E_{\text{th}}= N\, \langle \ve \rangle=2\,a^2\, N$. In the large $N$ limit, this equation amounts
to the condition $e= 2a^2$, i.e. the threshold energy is located exactly at the ($\beta = 0$)-line
drawn in Fig.~\ref{fig1}.

\begin{figure}
  \includegraphics[width=0.9\columnwidth]{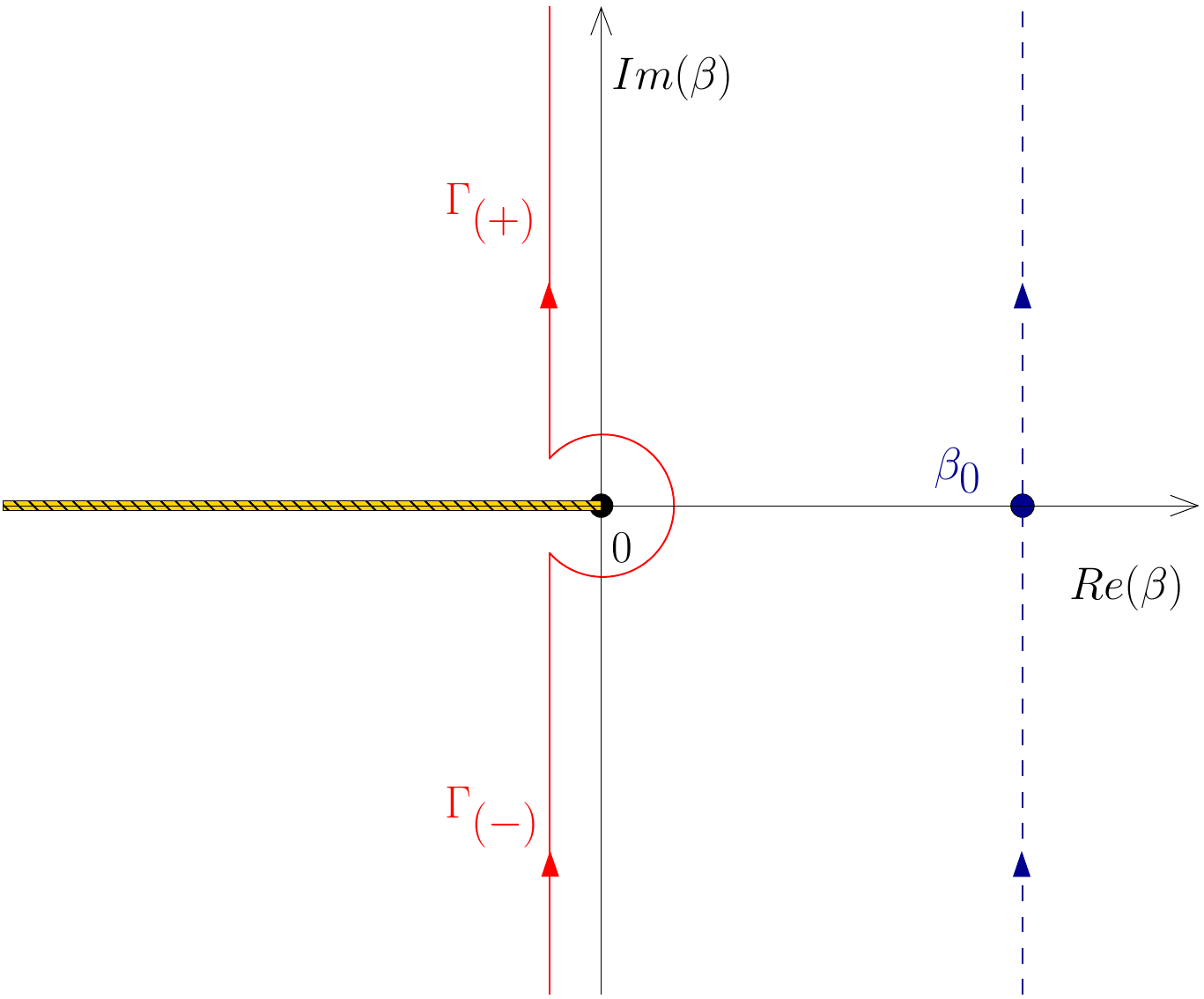}
  \caption{Analytic structure of the function
    \mycol{$z(\beta,\lambda)$} (see
      Eqs.~(\ref{eq:z-factorization}),(\ref{eq:z-onesite})) in
    the complex $\beta$ plane. 
    Slanted dashed band: branch cut on the negative real semiaxis. Dashed vertical
    (blue) line: Bromwich contour for the calculation of the partition
    function \mycol{$\mZ(\lambda,E)$} when
    \mycolor{$E<E_{\text{th}}$}, with $\beta_0$ indicating the
    location of the saddle-point. Continuous (red) line: Bromwich
    contour to compute $\mycol{\mZ(\lambda,E)}$ when
    \mycolor{$E>E_{\text{th}}$}. The critical energy \mycolor{$E_{\text{th}}\equiv E_{\text{th}}(A)$}, where the
    equivalence between ensembles breaks down, depends on the norm $A$.
    $\Gamma_{(+)}$ and
    $\Gamma_{(-)}$ are labels for contour pieces in the positive and
    negative imaginary semiplanes.}
\label{fig:Bromwich}
\end{figure}

\subsection{Analytic properties of the partition function}
\label{sec:2-B}

Before entering the main steps the microcanonical partition function
calculation, let us point out that this procedure is very similar to
the one employed in \cite{GM19} for a model of run-and-tumble
particles.

\gcol{Summarizing, the main steps of the calculation are the
  followings.} We start from the partition function $\mZ_N(\lambda,E)$
defined in Eq.~(\ref{eq:semipart}) and we first perform its Laplace
transform with respect to $E$, by \mycol{introducing its conjugate
  variable $\beta$, which is in general a complex number, i.e. $\beta
  \in \mathbb{C}$}:

\begin{align}
& \tilde\mZ_N(\lambda,\beta) = \int_0^\infty dE~e^{-\beta E}~\mZ_N(\lambda,E)  \nonumber \\
& = \int_0^\infty~d\varepsilon_1\ldots d\varepsilon_N~e^{-\beta \sum_{j=1}^N \ve_j}~\prod_{j=1}^Nf_\lambda(\ve_j)  \nonumber \\
& = \left[ \int_0^\infty d\ve~e^{-\beta \ve}~f_\lambda(\ve) \right]^N = \exp\left\lbrace N\log[z(\lambda,\beta)]\right\rbrace,
\label{eq:z-factorization}
\end{align}
where
\begin{align}
& z(\lambda,\beta) = \int_0^\infty d\ve~e^{-\beta \ve}~f_\lambda(\ve) \nonumber \\
& =  \sqrt{\pi}\frac{\lambda}{2\sqrt{\beta}}~\exp\left(\frac{\lambda^2}{4\beta}\right)~\text{Erfc}\left(\frac{\lambda}{2\sqrt{\beta}}\right),
\label{eq:z-onesite}
\end{align}
and
\be
\text{Erfc}\left(\frac{\lambda}{2\sqrt{\beta}}\right) = \frac{2}{\sqrt{\pi}}\int_{\lambda/(2\sqrt{\beta})}^\infty~e^{-t^2}~dt,
\ee
is the complementary error function defined in the complex $\beta$
plane, with a branch-cut on the negative real semiaxis (see
Fig.\ref{fig:Bromwich}). The original microcanonical partition function
$\mZ_N(\lambda,E)$ can be then recovered by computing the inverse
Laplace transform of \mycol{$\tilde\mZ_N(\lambda,\beta)$}:
\be
\mZ_N(\lambda,E) = \int_{\beta_0-i\infty}^{\beta_0+i\infty}
d\beta~\exp\left\lbrace N \left(\beta e + \log[z(\lambda,\beta)]\right)
\right\rbrace \, .
\label{eq:Zeta-gamma}
\ee
For large $N$, the integral in Eq.~(\ref{eq:Zeta-gamma})
can be evaluated using
a saddle-point approximation, where $\beta_0$ is the 
real positive solution of the following equation:
\be
e=\frac{E}{N} = - \frac{1}{z(\lambda,\beta)} \frac{\partial z(\lambda,\beta)}{\partial \beta} \,,
\label{eq:saddle-point-z}
\ee
If such a real positive value $\beta_0$ exists, it can be physically interpreted as an inverse temperature and
one could finally write
\be
\mZ_N(\lambda,E) ~\approx~e^{\beta_0 E + N \log[z(\lambda,\beta_0)]}.
\label{eq:saddle-point-b0}
\ee

What happens in the DNLSE model is that Eq.~(\ref{eq:saddle-point-z})
for $E<E_{\text{th}}$ admits a unique real positive solution and the
integration contour is shown by the dashed (blue) line in
Fig.\ref{fig:Bromwich}. Conversely, for $E>E_{\text{th}}$ a real
positive solution of Eq.~(\ref{eq:saddle-point-z}) does not exist, due
to the presence of the branch-cut of $z(\lambda,\beta)$ on the real
negative semiaxis. Accordingly, the equivalence between the canonical
and the microcanonical ensemble cannot hold for $E>E_{\text{th}}$.
Anyway, the calculation of $\mZ_N(\lambda,E)$ can be performed by
considering the analytic continuation of $z(\lambda,\beta)$ in the
complex $\beta$ plane and deforming the integration contour as shown
by the continuous (red) line in Fig.~\ref{fig:Bromwich}.  Following
similar large-deviation calculations, contained in a series of papers
~\cite{MEZ05,EMZ06,SEM14,GM19}, one can recover $\mZ_N(\lambda,E)$ for
$E>E_{\text{th}}$ making use of suitable expansions of
$z(\lambda,\beta)$ around the origin $\beta=0$.  
Consider evaluating
$\mZ_N(\lambda,E)$ for a given fixed scale $\Delta E =
E-E_{\text{th}}\sim N^\gamma $ of the \emph{excess} energy 
where the exponent $\gamma$ can be chosen depending on
which scale we want to probe the system.
One then needs
to retain the leading terms only up to a given scale in the expansion
of $z(\lambda,\beta)$. For instance, as explicitly reported in
~\cite{GM19}, one can evaluate $\mZ_N(\lambda,E)$ at the energy scale
$\Delta E = E-E_{\text{th}}\sim N^{1/2}$ (Gaussian regime) by
expanding $z(\lambda,\beta)$ around $\beta=0$ up to the order
$\beta\sim N^{-1/2}$, thus obtaining the almost trivial result
\be
\mZ_N(\lambda,E) = \frac{1}{\sigma\sqrt{2\pi N}}\exp\left[ - \frac{(E - E_{\text{th}})^2}{2\sigma^2 N} \right]\, ,
\label{eq:ZE-Gaussian}
\ee
which is a straightforward consequence of the Central Limit Theorem,
the dependence on $\lambda$ coming through $\sigma$ (see
  Eq.~(\ref{eq:moments-1-2})). On the other hand, if one aims at
evaluating $\mZ_N(\lambda,E)$ at the order $\Delta E =
E-E_{\text{th}}\sim N$, which in~\cite{GM19} is called the
\emph{extreme large-deviation} regime, one has to retain consistently
terms of the expansion of $z(\lambda,\beta)$ up to the order $\beta
\sim 1/N$.  In this case one obtains:
\be
\mZ_N(\lambda,E) \sim \exp\left( - \sqrt{E-E_{\text{th}}} \right).
\label{eq:ZE-asymptotic}
\ee
As discussed in Sec.~\ref{sec:3-B}, the Gaussian regime and the extreme
large-deviation regime correspond to a delocalized and to a localized
phase, respectively. Therefore, in order to understand whether the
crossover between these two phases occurs as a real thermodynamic
phase transition, one has to identify an intermediate
\emph{matching} regime, which can be heuristically singled out
by the following condition
\be
\frac{(E - E_{\text{th}})^2}{2\sigma^2 N} \approx \sqrt{E-E_{\text{th}}}\,,
\label{eq:match}
\ee
\mycol{allowing to recognize the intermediate scale
\be
E - E_{\text{th}} \sim N^{2/3}.
\ee
}

\subsection{\emph{Matching} regime: the non-analytic contribution}
\label{sec:2-C}

The main result of this section is the proof that in the matching
regime the partition function splits into the sum of a Gaussian
contribution and of a \mycol{non-analytic one, denoted by
 $\mC(\lambda,\zeta)$}:
\begin{align}
\mZ_N(\lambda,E) = \frac{1}{\sigma\sqrt{2\pi N}} \exp{\left(- N^{1/3} \frac{\zeta^2}{2\sigma^2}\right)} + \mycol{\mC(\lambda,\zeta)} \, ,
\label{eq:Z-splitting}
\end{align}
where
\begin{align}
\zeta = \frac{E-E_{\text{th}}}{N^{2/3}}
\label{eq:scaling-variable}
\end{align}
is the scaling variable suggested by the \emph{matching} condition in
Eq.~(\ref{eq:match}).  The first term on the r.h.s. of
Eq.~(\ref{eq:Z-splitting}) comes from the straight part of the
deformed contour in Fig.~\ref{fig:Bromwich} (continuous red line),
while the \mycol{non-analytic} contribution,
\mycol{$\mC(\lambda,\zeta)$}, is due to the non-analyticity at the
branch-cut along the negative real axis.\\

Since for $E>E_{\text{th}}$ the saddle-point condition in
Eq.~(\ref{eq:saddle-point-z}) has no real solution, we have to
consider the analytic prolongation of $z(\lambda,\beta)$ in the
complex $\beta$ plane and to evaluate its expansion around the origin
$\beta=0$ separately along the upper and lower branch cut, i.e.  for
$\Re(\beta)<0$:
\bea
\lim_{\delta\rightarrow 0} z(\lambda,\beta+i\delta) &=& z(\lambda,\beta+i0^{+}) \nonumber \\
\lim_{\delta\rightarrow 0} z(\lambda,\beta-i\delta) &=& z(\lambda,\beta+i0^{-}).
\eea
The expansion around $\beta=0$ yields the expressions
\begin{align}
  & z(\lambda,\beta + i0^{+} ) = \nonumber \\
  & 1 - \langle \ve \rangle \beta + \frac{1}{2} \langle \ve^2 \rangle \beta^2 +\ldots+
\sqrt{\frac{2\pi}{\langle \ve \rangle \beta}}~\exp\left(\frac{1}{2\langle \ve \rangle\beta}\right) \nonumber \\
& z(\lambda,\beta + i0^{-} ) = 1 - \langle \ve \rangle \beta + \frac{1}{2} \langle \ve^2 \rangle \beta^2+\ldots,
\label{eq:expansion-g-beta}
\end{align}
where $\langle \ve \rangle$  and $\langle \ve^2 \rangle$ are defined in Eq.~(\ref{eq:moments-1-2}) in terms of $\lambda$.
Accordingly, the expansion of the local free energy reads:
\begin{align}
  & \log[z(\lambda,\beta+i0^{+})] = \nonumber \\
  & - \langle \ve \rangle~\beta +\frac{1}{2}\sigma^2~\beta^2  + \mO(\beta^3)
+ \ldots + \sqrt{\frac{2\pi}{\langle \ve \rangle \beta}}~\exp\left(\frac{1}{2\langle \ve \rangle\beta}\right) \nonumber \\
& \log[z(\lambda,\beta+i0^{-})] = - \langle \ve \rangle~\beta +\frac{1}{2}\sigma^2~\beta^2  + \mO(\beta^3),
\label{eq:expansion-log-g-beta}
\end{align}
where $\sigma$ is defined in Eq.~(\ref{eq:moments-1-2}) in terms of
$\lambda$.  The non-analiticity of $z(\lambda,\beta)$ at the cut,
clearly expressed by the difference between the first and the second
line of Eq.~(\ref{eq:expansion-log-g-beta}), suggests to evaluate the
contour integral which defines $\mZ_N(\lambda,E)$ [see
  Eq.~(\ref{eq:Zeta-gamma})] by splitting it in two parts:
\be
\mZ_N(\lambda,E) = \mI^{(+)}(\lambda,E) + \mI^{(-)}(\lambda,E) \, .
\ee
The two terms $\mI^{(+)}(\lambda,E)$ and $\mI^{(-)}(\lambda,E)$ in the
equation above can be evaluated by introducing explicitly the
expansions of Eq.~(\ref{eq:expansion-log-g-beta}) into the integral of
Eq.~(\ref{eq:Zeta-gamma}). Recalling then that $E_{\text{th}} =
N\langle \ve \rangle $ one obtains
\begin{widetext}
\bea
\mI^{(+)}(\lambda,E) &=& \int_{\Gamma^{(+)}}\frac{d\beta}{2\pi i}~\exp\left\lbrace \beta (E - E_{\text{th}}) + \frac{N}{2}\sigma^2\beta^2 + N \mO(\beta^3) +
N \sqrt{\frac{2\pi}{\langle \ve \rangle \beta}}~\exp\left(\frac{1}{2\langle \ve \rangle\beta}\right) \right\rbrace \nonumber \\
\mI^{(-)}(\lambda,E) &=& \int_{\Gamma^{(-)}}\frac{d\beta}{2\pi i}~\exp\left\lbrace \beta (E - E_{\text{th}}) + \frac{N}{2}\sigma^2\beta^2 + N \mO(\beta^3) \right\rbrace \,, \nonumber \\
\eea
where the integration paths $\Gamma^{(+)}$ and $\Gamma^{(-)}$ are
those shown in Fig.\ref{fig:Bromwich}.  A better interpretation of
this result can be obtained by introducing the scaling variable
$\zeta$ defined in Eq.~(\ref{eq:scaling-variable}):
\bea
\mI^{(+)}(\lambda,\zeta) &=& \int_{\Gamma^{(+)}}\frac{d\beta}{2\pi i}~\exp\left\lbrace \beta~\zeta~N^{2/3} + \frac{N}{2}\sigma^2\beta^2 + N \mO(\beta^3) + N \sqrt{\frac{2\pi}{\langle \ve \rangle \beta}}~\exp\left(\frac{1}{2\langle \ve \rangle\beta}\right) \right\rbrace \nonumber \\
\mI^{(-)}(\lambda,\zeta) &=& \int_{\Gamma^{(-)}}\frac{d\beta}{2\pi i}~\exp\left\lbrace \beta~\zeta~N^{2/3} + \frac{N}{2}\sigma^2\beta^2 + N \mO(\beta^3) \right\rbrace. \nonumber \\
\label{eq:integrals-Ipm}
\eea
Since $\Re(\beta)<0$, for asymptotically small values of $\beta$
one has that the non-analytic term $\exp[1/(2\langle \ve \rangle\beta)]$ is
exponentially small and, at  leading order in $N$, we can write
\begin{align}
\exp \left[ N \sqrt{\frac{2\pi}{\langle \ve \rangle
      \beta}}~\exp\left(\frac{1}{2\langle \ve \rangle\beta}\right)
  \right] \approx 1 + N \sqrt{\frac{2\pi}{\langle \ve \rangle
    \beta}}~\exp\left(\frac{1}{2\langle \ve \rangle\beta}\right).
\end{align}
\end{widetext}
By substituting the above expansion into the integral $\mI^{(+)}(\lambda,\zeta)$ we obtain
\begin{align}
& \mI^{(+)}(\lambda,\zeta) = \int_{\Gamma^{(+)}} \frac{d\beta}{2\pi i}~e^{\beta \zeta N^{2/3} + \frac{N}{2}\sigma^2\beta^2 + N \mO(\beta^3)} + \nonumber \\
& + N \sqrt{\frac{2\pi}{\langle \ve \rangle \beta}}  \int_{\Gamma^{(+)}} \frac{d\beta}{2\pi i}~
e^{\beta \zeta N^{2/3} + \frac{N}{2}\sigma^2\beta^2 + N \mO(\beta^3) + \frac{1}{2\langle \ve \rangle\beta}}.
\label{eq:Iplus-exp}
\end{align}
A more refined way to implement the matching argument at the end
of~\ref{sec:2-B} is to set the scale of the (complex) neighbourhood of
$\beta=0$ in such a way that the analytic and the non-analytic terms
in the expansion of $\log[z(\lambda,\beta)]$ are of the same order in
$N$: \gcol{in fact, these terms account respectively for the
  homogeneous and the localized phase}.

Since we want  to evaluate the
integrals in Eq.~(\ref{eq:Iplus-exp}) by means of a saddle-point
approximation,  we have to chose the proper scale of
$\beta$. This task can be accomplished by first introducing the change of variable to $\hat{\beta} =
N^{\gamma} \beta$ in the integral of Eq.~(\ref{eq:Iplus-exp}) and then by
fixing the value of the exponent $\gamma$ in such a way to single out the same
power of $N$ in each term of the following expression:
\begin{align}
\hat{\beta} \zeta~N^{2/3-\gamma} + \frac{N^{1-2\gamma}}{2}\sigma^2\hat{\beta}^2 + \frac{N^{\gamma}}{2\langle \ve \rangle \hat{\beta}}.
\label{eq:anomalous-matching-0}
\end{align}
This is condition holds if there is a value of $\gamma$ such that
\be
2/3-\gamma = 1-2\gamma = \gamma,
\label{eq:anomalous-matching}
\ee
thus yielding  $\gamma=1/3$.
The integrals in Eq.~(\ref{eq:Iplus-exp}) can be rewritten in the following form
\begin{align}
& \mI^{(+)}(\lambda,\zeta) = \frac{1}{N^{1/3}} \int_{\Gamma^{(+)}} \frac{d\beta}{2\pi i}~e^{N^{1/3}[\beta \zeta+ \frac{1}{2}\sigma^2\beta^2]} + \nonumber \\
& N \sqrt{\frac{2\pi}{\langle \ve \rangle}}  \int_{\Gamma^{(+)}} \frac{d\beta}{2\pi i}~\frac{1}{\sqrt{N^{1/3}\beta}}
e^{N^{1/3} \left[ \beta \zeta + \frac{1}{2}\sigma^2\beta^2 + \frac{1}{2\langle \ve \rangle\beta}\right]},
\end{align}
where the integration variable $\beta$ has been rescaled by a factor $N^{1/3}$.
Similarly, the integral in the negative imaginary semiplane turns to the expression
\be
\mI^{(-)}(\lambda,\zeta) = \frac{1}{N^{1/3}} \int_{\Gamma^{(-)}} \frac{d\beta}{2\pi i}~e^{N^{1/3}[\beta \zeta+ \frac{1}{2}\sigma^2\beta^2]}\, .
\ee
By summing these two contributions one finally obtains the partition
function $\mZ_N(\lambda,E)$. It is convenient to change the variable
$E$ to the scaled variable $\zeta=(E-E_{\rm th})/N^{2/3}$
in $\mZ_N(\lambda,E)$, and with a slight abuse of notation, we
will continue to denote the partition function as $\mZ_N(\lambda,\zeta)$.
We then obtain
\bea
\mZ_N(\lambda,\zeta) &=& \mI^{(+)}(\lambda,\zeta) + \mI^{(-)}(\lambda,\zeta) = \nonumber \\
&=& \frac{1}{N^{1/3}} \int_{-i\infty}^{i\infty} \frac{d\beta}{2\pi i}~e^{N^{1/3}[\beta \zeta+ \frac{1}{2}\sigma^2\beta^2]} + \mycol{\mC(\lambda,\zeta)}, \nonumber \\
&=& \frac{1}{\sigma\sqrt{2\pi N}} e^{- N^{1/3} \frac{\zeta^2}{2\sigma^2}} + \mycol{\mC(\lambda,\zeta)}
\label{eq:instantona}
\eea
where the \mycol{non-analytic} contribution to the partition function is
\be \mycol{\mC(\lambda,\zeta)} = N \sqrt{\frac{2\pi}{\langle \ve \rangle}}
\int_{\Gamma^{(+)}} \frac{d\beta}{2\pi
  i}~\frac{1}{\sqrt{N^{1/3}\beta}} e^{N^{1/3} F_{\zeta}(\lambda,\beta)}\, ,
\label{eq:instanton}
\ee
with
\be
F_{\zeta}(\lambda,\beta) =  \beta \zeta + \frac{1}{2}\sigma^2\beta^2 + \frac{1}{2\langle \ve \rangle\beta} \, .
\label{eq:Fz}
\ee
Note that the dependence of $F_{\zeta}(\lambda,\beta)$ on $\lambda$
comes from the fact that both $\langle \ve \rangle$ and $\sigma^2$
appearing in Eq. (\ref{eq:Fz}) are functions of $\lambda$ [see
Eq.~(\ref{eq:moments-1-2})].\\

The decomposition of the partition function in the matching regime as
the sum of two contributions is the main result of this section. It
remains to perform the explicit calculation of the integral
$\mycol{\mC(\lambda,\zeta)}$.\\

This task can be accomplished by following the same procedure reported
in~\cite{GM19}: the key point of the calculation amounts to finding
the solution $\beta^*(\zeta)$ of the saddle-point equation:
\be
\frac{\partial F_{\zeta} (\lambda,\beta)}{\partial \beta} = 0\, .
\label{eq:saddle-point-Fz}
\ee
Details of the calculations are illustrated in Appendices \ref{sec:app-A} and \ref{sec:app-B}. Here we just
provide the final result:
\be
\int_{\Gamma^{(+)}}\frac{d\beta}{2\pi i}~\frac{1}{\sqrt{N^{1/3}\beta}}~e^{N^{1/3}F_{\zeta}(\lambda,\beta)} \approx e^{- N^{1/3} \chi(\zeta)} \,,
\label{eq:chi-def}
\ee
where the explicit form of $\chi(\zeta)$ is discussed in
Appendix~\ref{sec:app-A}.
\mycol{Its asymptotic behaviours are:}
\be
\chi(\zeta) = \begin{cases}
\frac{3}{2}\left(\frac{\sigma}{\langle \varepsilon \rangle}\right)^{2/3}
\qquad\qquad\qquad\qquad \zeta\rightarrow \zeta_l \\ \\
\sqrt{\frac{2}{\langle \varepsilon \rangle}} \sqrt{\zeta}-\frac{\sigma^2}{4\langle\varepsilon\rangle}\frac{1}{\zeta}+\mO\left(\frac{1}{\zeta^{5/2}}\right),
\quad \zeta\gg 1 \end{cases},
\label{eq:chi-asymptotics-intro}
\ee
where $\zeta_l$ is the spinodal point for the localized phase, that is the
smallest value of $\zeta$ for which the saddle-point equation Eq.~(\ref{eq:saddle-point-Fz})
admits a real solution, namely
\be
\zeta_l = \frac{3}{2}~\left(\frac{\sigma^4}{\langle \varepsilon \rangle}\right)^{1/3}.
\label{eq:spinodal-zl}
\ee


\section{The  first-order phase transition from a thermalized phase to localization}
\label{sec:3}

\begin{figure}
\includegraphics[width=0.9\columnwidth]{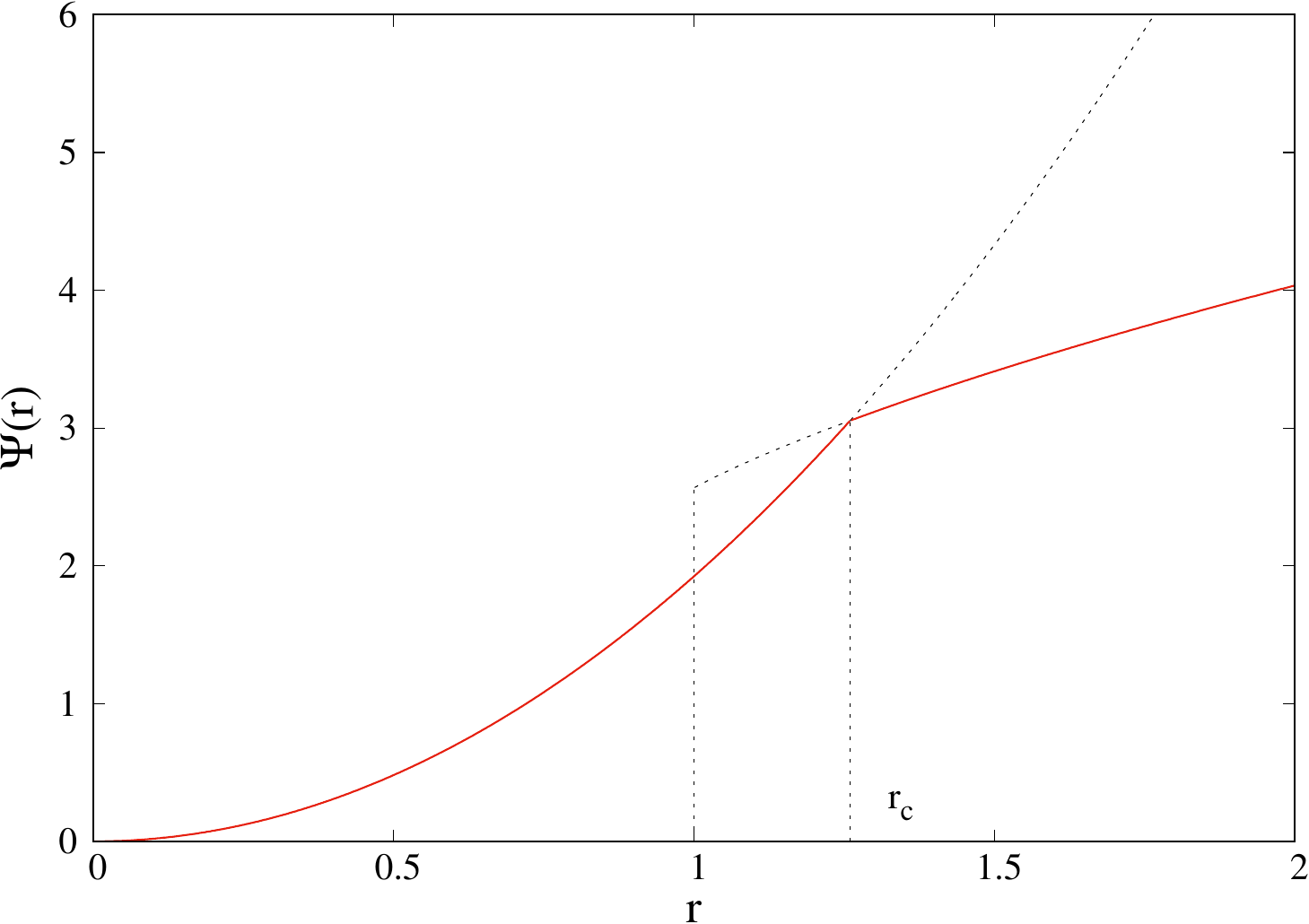}
\caption{Continuous (red) line: behaviour of the sub-leading
  contribution $\Psi(r)$ (see Eq.~(\ref{eq:rate-func})) to the
  microcanonical entropy as a function of the adimensional variable $r
  = \zeta/\zeta_l$, with $\zeta_l$ defined in
  Eq.~(\ref{eq:spinodal-zl}). Making use of this rescaled variable
  $r$, the spinodal point is located at $r=1$.  The first order
  transition is located at $r_c= \zeta_c/\zeta_l\approx 2^{1/3}$,
  where $d\Psi(r)/d r$ is discontinuous: the value of $\zeta_c$ is
  determined by the argument in Appendix~\ref{sec:app-D}. The dashed lines draw
  the function $\chi(r)$ (see
    Eqs.~(\ref{eq:chi-def}),(\ref{eq:chi-asymptotics-intro})) for
  $r<r_c$ and $r^2/(2\sigma^2)$ for $r>r_c$.}
 \label{fig:transition}
\end{figure}
\subsection{The microcanonical entropy in the matching regime}
\label{sec:3-A}

We are now in the position of retrieving the microcanonical partition
function \mycol{by computing the inverse Laplace transform of
  Eq.~(\ref{eq:invL-A}):}
\begin{align}
& \mycol{\Omega_N(A,E)} = \frac{1}{2\pi i}\int_{\lambda_0-i\infty}^{\lambda_0+i\infty} d\lambda~e^{\lambda A}~\tilde{\Omega}_N(\lambda,E) \nonumber \\
& = \frac{e^{N\log(\pi)}}{2\pi i}\int_{\lambda_0-i\infty}^{\lambda_0+i\infty} d\lambda~e^{N [a\lambda-\log(\lambda)]}~\mZ_N(\lambda,E),
\label{eq:Laplace-inverse-mu}
\end{align}
where the final expression on the r.h.s. of
Eq.~(\ref{eq:Laplace-inverse-mu}) is obtained thanks to
Eq.~(\ref{eq:canonical-2}). In order to point out the presence of a
phase transition, we are interested to obtain an analytic estimate of
$\mycol{\Omega_N(A,\zeta)}$ in the matching regime, where
$\mZ_N(\lambda,\zeta)$ is given by Eqs.~(\ref{eq:instantona}) and
Eq.~(\ref{eq:instanton}) and $\zeta = (E - E_{\text{th}})/N^{2/3}
\approx \mO(1)$:
\begin{align}
  & \mycol{\Omega_N(A,\zeta)} = \nonumber \\
  & \frac{e^{N\log(\pi)}}{2\pi i}\int_{\lambda_0-i\infty}^{\lambda_0+i\infty} d\lambda~e^{N [a\lambda-\log(\lambda)]}~
\left[ e^{- N^{1/3} \chi(\zeta)} + e^{- N^{1/3}\zeta^2/(2\sigma^2)} \right].
\label{eq:Laplace-inverse-mu-matching}
\end{align}
In the thermodynamic limit, $N \to \infty$, the leading contribution
to the integral on the r.h.s. of
Eq.~(\ref{eq:Laplace-inverse-mu-matching}) is given by the term $N
[a\lambda-\log(\lambda)]$, which determines the value of $\lambda_0$
as the solution of the saddle-point equation
\be
\frac{\partial}{\partial \lambda} [a\lambda-\log(\lambda)]=0 \qquad \longrightarrow \qquad \lambda_0 = 1/a \,.
\label{spc}
\ee
The complete expression of $\mycol{\Omega_N(A,\zeta)}$ is thus
obtained by replacing the multiplier $\lambda$ with its actual value
$\lambda_0 = 1/a$. Accordingly, in the matching regime the complete
partition function reads
\be
\mycol{\Omega_N(A,\zeta)}~\approx~e^{N\left[ 1 + \log(\pi a) \right]}~\left[ e^{- N^{1/3} \chi(\zeta)} + e^{- N^{1/3}\zeta^2/(2\sigma^2)} \right],
\label{eq:micro-matching}
\ee
\gcol{where we point out that, by plugging $\lambda=1/a$ into
  Eq.~(\ref{eq:moments-1-2}), one has $\sigma^2 = 20\, a^4$ and
  $\langle \ve \rangle = 2\, a^2$}. Note that we need $\langle \ve
\rangle$ to evaluate $\chi(\zeta)$ in
Eq.~(\ref{eq:chi-asymptotics-intro}).  This result provides us with
the expression for the microcanonical entropy at leading and
sub-leading order in the limit of large $N$:
\be
S_N(A,\zeta) = N\left[ 1 + \log(\pi a) \right] - N^{1/3} \Psi(\zeta),
\label{eq:rate-func-all}
\ee
where
\be
\Psi(\zeta) = \text{inf}_{\zeta}\Big\{\chi(\zeta),  \zeta^2/(2\sigma^2)\Big\}\, ,
\label{eq:rate-func}
\ee
The leading term in $S_N(A,\zeta)$ is extensive in $N$ and represents
the contribution of the bulk of the system at infinite temperature, i.e. the
\emph{background entropy},
\be
S_N^{\text{back}}(a) = N\left[ 1 + \log(\pi a) \right] = N s^{\text{back}}(a) .
\ee
Note that this contribution does not depend on $\zeta$, i.e. on the excess energy $\Delta E
= E - E_{\text{th}}$ contained in the condensate, which appears  
exclusively only in the sub-leading term in $\Psi(\zeta)$.
Accordingly, we can identify the critical value $\zeta_c$ for
localization as the one where the sub-leading contribution to the entropy
of the localized and that of the delocalized phase have identical
magnitude, i.e., by the following matching condition
\be
\chi(\zeta_c) = \zeta_c^2/(2\sigma^2)\, .
\label{eq:crit-val}
\ee
The function $\Psi(\zeta)$ is shown in Fig.\ref{fig:transition}: more
precisely, we have decided to draw it as a function of the rescaled
variable $r = \zeta/\zeta_l$. {\color{black} From the argument
  discussed in App.~\ref{sec:app-D} we find that the critical value of
  $r$, which does not depend on the parameters of individual energy
  distributions on lattice sites, is
\be
r_c = \frac{\zeta_c}{\zeta_l}=2^{1/3}.
\label{eq:crit-spin-ratio}
\ee
}
Since the derivative of $\Psi(\zeta)$ is
discontinuous at $\zeta_c$, we can conclude that we are facing a
first--order phase transition, from a thermalized phase to a localized
one.\\

\subsection{The order parameter: participation ratio}
\label{sec:3-B}

The localization transition can be further characterized by introducing
the participation ratio of the energy per site
\begin{align}
Y_2 = \left\langle \frac{\sum_{j=1}^N \ve_j^2}{\left(\sum_{j=1}^N \ve_j\right)^2}\right\rangle_{A,E},
\label{eq:part-ratio}
\end{align}
as a suitable order parameter, where the angular brackets denote the
microcanonical equilibrium average. Indeed, if the total energy $E=e
N$ is uniformly distributed over all sites, i.e.  $\varepsilon_j \sim
e $ for each $j$, then the numerator in Eq.~(\ref{eq:part-ratio})
scales as $N$ and the denominator scales as $N^2$. Consequently $Y_2 =
\mathcal{O}(N^{-1})$.  In contrast, when an extensive amount of energy
is localized at a single site (i.e. in the presence of a condensate),
the numerator scales as $N^2$ while the denominator still scales as
$N^2$. Therefore, $Y_2= \mathcal{O}(1)$ in the large $N$ limit. Hence,
the quantity $Y_2$ is a good observable to detect the localization
transition.

For our purposes it is enough to analyze the behavior of $Y_2$ in the
proximity of the threshold energy, i.e., at $E \sim E_{\text{th}} =
N\langle \varepsilon \rangle$ (see Sec.~\ref{sec:2-A}), where, as
discussed in the previous section, the transition point is located.
In this regime the equilibrium joint probability distribution of local
energies has, within the microcanonical ensemble, the following
expression
\begin{align}
P(\ve_1,\ldots,\ve_N) = \frac{e^{N s^{\text{back}}(a)}}{\mycol{\Omega_N(A,E)}}\prod_{j=1}^N~f_a(\ve_j)~\delta\left( E - \sum_{j=1}^N \ve_j \right) \,,
\label{eq:prob-conf-micro}
\end{align}
where
\begin{align}
  & \mycol{\Omega_N(A,E)} = \nonumber \\
  & e^{N s^{\text{back}}(a) } \int_0^\infty \prod_{j=1}^N~\left[d\ve_j~f_a(\ve_j)~\right]~\delta\left( E - \sum_{j=1}^N \ve_j \right)
\label{eq:Z-one-constraint}
\end{align}
is the microcanonical partition function. In fact, as shown in the
previous section [see the saddle-point condition in Eq.~(\ref{spc})],
$\mycol{\Omega_N(A,E)}$ in the matching regime can be written as a
function of the probability distributions $f_a(\ve_j)$ of
i.i.d. energy variables, where \mycol{the label} $\lambda$ appearing
in Eq.~(\ref{eq:fx}) can be replaced by the label $a$, \gcol{since by
  means of the saddle-point condition in Eq.~(\ref{spc}) we have fixed
  $\lambda=1/a$}, i.e., the
inverse of the solution of the saddle-point condition. By plugging the
expression of $P(\ve_1,\ldots,\ve_N)$ written in
Eq.~(\ref{eq:prob-conf-micro}) into the definition of $Y_2$ given in
Eq.~(\ref{eq:part-ratio}), we obtain:
\begin{widetext}
\begin{align}
Y_2(E) &= \frac{e^{N s^{\text{back}}(a)} }{\mycol{\Omega_N(A,E)}} \int_0^\infty d\ve_1\ldots d\ve_N~
\prod_{j=1}^N~f_a(\ve_j)~\frac{\sum_{j=1}^N \ve_j^2}{\left(\sum_{j=1}^N \ve_j\right)^2}~\delta\left( E - \sum_{j=1}^N \ve_j \right)\nonumber \\
&= \frac{N e^{s^{\text{back}}(a)} }{E^2} \int_0^\infty d\ve~\ve^2~f_a(\ve)~
\frac{\left[e^{(N-1) s^{\text{back}}(a)}~\int d\ve_2\ldots d\ve_N \prod_{j=2}^N f_a(\ve_j) \delta\left( E-\ve-\sum_{j=2}^N \ve_i\right)\right]}{\mycol{\Omega_N(A,E)}} \nonumber \\
&= \frac{N e^{s^{\text{back}}(a)}}{E^2} \int_0^\infty d\ve~\ve^2~f_a(\ve)~\frac{\Omega_{N-1}(A,E-\ve)}{\mycol{\Omega_N(A,E)}} \,,
\label{eq:participation}
\end{align}
\end{widetext}
where the term
\be
e^{s^{\text{back}}(a)} f_a(\ve) \frac{\mycol{\Omega_{N-1}(A,E-\ve)}}{\mycol{\Omega_{N}(A,E)}} \equiv \rho(\ve)
\label{eq:marginal-ene}
\ee
is the normalized marginal distribution  of the energy $\ve$
on a single site
~\footnote{The normalization of $\rho(\ve)$ can be verified 
from a calculation  analogous to the one in Eq.~(\ref{eq:participation}) for the microcanonical average of 
the constant function $f(E)=1$.}.  
   Accordingly, the participation ratio takes the compact form
 \bea
Y_2(E) &=& \frac{N  }{E^2} \int_0^\infty d\ve~\ve^2~\rho(\ve) \nonumber \\
&=&  \frac{N}{E^2}~\langle \ve^2 \rangle \,,
\label{eq:participation2}
\eea

As a first step in the study of the behavior of the order parameter
$Y_2$, \gcol{ we observe that the condition $E^2 \sim N^2$ implies that
  Eq.~(\ref{eq:participation2}) reads in practice as $Y_2(E) \sim
  \langle \ve^2 \rangle/N$.

  Let us first consider the estimate of $Y_2$ in the case when
  $E<E_{th}$. Since the total energy is below the threshold,
  $\Omega_N(A,E)$ can be safely computed by means of a saddle point
  approximation, yielding
  \begin{align}
    \Omega_N(A,E) \approx e^{\beta_0 E + N \log[z(a,\beta_0)]}.
  \end{align}

}
%
%
It is then crucial the calculation of $\mycol{\Omega_{N-1}(A,E-\ve)}$
in Eq.~(\ref{eq:marginal-ene}), which, for large values of $N$, can be
approximated in a completely harmless way with
$\mycol{\Omega_N(A,E-\ve)}$. Since the value of the energy $\ve$ on a
single site can be {\it at most} $\ve=E$, we have that the domain of
the variable $y=E-\ve$ is $y\in [0,E]$. Hence the integral which
defines $\mycol{\Omega_N(A,y=E-\ve)}$ through its inverse Laplace
transform can be computed by a saddle-point approximation \gcol{for
  all values of $y$, since we always have $y=E-\varepsilon < E_{th}$}:
\begin{align}
  \mycol{\Omega_N(A,E-\ve)} & = \int_{\beta_0-i\infty}^{\beta_0+i\infty} d\beta~e^{\beta (E-\ve) + N\log[z(a,\beta)]} \nonumber \\
  & \approx e^{\beta_0 (E-\ve) + N\log[z(a,\beta_0)]} \nonumber \\
\end{align}
so that
\begin{align}
\frac{\mycol{\Omega_{N-1}(A,E-\ve)}}{\mycol{\Omega_N(A,E)}} = e^{-\beta_0 \ve}~\Longrightarrow~\frac{\langle \ve^2 \rangle}{N} \sim \frac{1}{N}.
\label{eq:inv-Laplace-Y2}
\end{align}
Accordingly, for $E<E_{\text{th}}$ the participation ratio vanishes as
$1/N$ for large $N$.  This confirms that in the thermalized phase
close to $E_{\textrm{th}}$ localization of energy is absent, as
expected. For $E>E_{\textrm{th}}$ we cannot rely anymore on the
saddle-point approximation for $\beta$ to compute the integral in
Eq.~(\ref{eq:inv-Laplace-Y2}). Still, from the study of
$\mycol{\Omega_N(A,E)}$ we known that at the scale $E-E_{\text{th}}
\sim N^{1/2}$ the partition function has a Gaussian shape. Therefore,
we have that for values of $E$ up to the scale $E-E_{\text{th}} \sim
N^{1/2}$ the marginal probability distribution of energy scales as
\be
\mycolor{ \rho(\ve)}  \approx f_a(\ve) \frac{\exp[-(\Delta E-\ve)^2/(2\sigma^2 N)]}{\exp[-(\Delta E)^2/(2\sigma^2 N)]}\, ,
\label{eq:marginal-gauss-p}
\ee
where $\Delta E = E-E_{\text{th}}$. By expanding the
square in the exponential in the numerator  we obtain a term which simplifies with the denominator
and we are left with the expression:
\be
\mycolor{\rho(\ve)} \approx f_a(\ve)~\exp\left[-\frac{\ve^2}{2\sigma^2 N}+\frac{\ve \Delta E}{\sigma^2 N} \right] \, .
\label{eq:marginal-gauss}
\ee
Recalling now that $\Delta E \sim \sqrt{N}$, we can estimate the participation ratio by the relation
\be
Y_2 \approx  \frac{1}{N} \int_0^\infty d\ve~\ve^2~f_a(\ve)~
\exp\left[-\frac{\ve^2}{2\sigma^2 N}+\frac{\ve}{\sigma^2 \sqrt{N}} \right]\, .
\ee
In the limit of large $N$ this integral can be approximated as
\be
Y_2 \approx  \frac{1}{N} \int_0^\infty d\ve~\ve^2~f_a(\ve) \sim \frac{1}{N}\, ,
\ee
Therefore for $E>E_{\text{th}}$ and $E-E_{\text{th}}\sim \sqrt{N}$ the
participation ratio vanishes asymptotically with $N$. More precisely,
the system is still in a delocalized phase, although the decay of
$\rho(\ve)$ is not monotonic (see also Fig.~\ref{fig:bump}).  With the
same argument it can be shown that the same scaling of $Y_2$ persists
even for $E-E_{\text{th}}\sim N^{2/3}$, i.e. in the matching regime.
Altogether, we conclude that the system is delocalized up to $E_c$,
the critical value of the total energy where the first derivative of
the function $\Psi(\zeta)$ in Eq.~(\ref{eq:rate-func-all}) exhibits
the discontinuity.  According to Eq.~(\ref{eq:scaling-variable}),
$E_c$ reads as:
\be
E_c = E_{\text{th}} + N^{2/3} \zeta_c,
\label{eq:critical-E}
\ee
where the $\zeta_c$ does not depend on $N$ and is determined by the
matching condition in Eq.~(\ref{eq:crit-val}), see also
Appendix~\ref{sec:app-D}).

The situation is different for the case of \emph{extreme} large
deviations of the total energy, i.e., following the terminology
of~\cite{GM19}, for $E-E_{\text{th}} \sim N$.  Also in this case the
marginal distribution $\rho(\ve)$ exhibits a bump (see
Fig.~\ref{fig:bump}). But in this case the whole $\rho(\ve)$ is
dominated in the large $N$ limit by the contribution of the bump,
which, for fluctuations of order $\Delta E -\ve \sim N^{1/2}$ around
the bump center, reads as
\be
\rho(\ve) \approx \rho_{\text{bump}}(\ve) = f_a(\ve) ~\frac{1}{N
  f_a(\Delta E)}~\frac{1}{\sqrt{N}} e^{-(\Delta E-\ve)^2/(2\sigma^2
  N)}, \ee
where we have used the fact that in the extreme large-deviation
regime the whole partition function is identical to $N$ times the
distribution of the single variable (see \cite{MEZ05,SEM14}), in
formulae
\be
\mycol{\Omega_N(A,E)} \approx N~f_a(E-E_{\text{th}}).
\ee
Since we are interested to the estimate of
$\rho_{\text{bump}}(\ve)$ for $\ve\sim \Delta E$, we have
$f_a(\ve)/f_a(\Delta E)=\mO(1)$ in the large $N$ limit. Therefore
\begin{align}
& \rho_{\text{bump}}(\ve) ~\approx~ \frac{1}{N^{3/2}}~\exp[-(\Delta E-\ve)^2/(2\sigma^2 N)]~\nonumber \\
& \approx~\frac{1}{N}~\delta(\Delta E -\ve).
\label{eq:Gauss-bump}
\end{align}
By recalling that $\Delta E \sim N $, we finally obtain
\be
\int_0^\infty d\ve~\ve^2~\rho_{\text{bump}}(\ve) = \frac{(\Delta E)^2}{N} \sim N,
\ee
so that in this limit $Y_2$ is finite and independent of $N$.  These
features signal the presence of the localized phase. We point out that
localization of energy is on a single site, as the ratio between the
width of the bump, of order $N^{1/2}$, and its position $\Delta E \sim
N$, vanishes asymptotically.\\

\mycol{Overall, the scenario can be summarized according to the
  following scheme, where the critical value of the energy $E_c$ is
  the one defined in Eq.~(\ref{eq:critical-E}):}
\begin{itemize}
\item[(A)] $E<E_{\text{th}}$: $\mycolor{\rho(\ve)}$ decays
  monotonically at large $\ve$ and the participation ratio decays
  asymptotically as
  \be
  Y_2(E) \sim \frac{1}{N} \, .
  \ee
\item[(B)] $E_{\text{th}}<E<E_c$: the decay of $\mycolor{\rho(\ve)}$ at
  large values of $\ve$ is non monotonic and the formation of a
  secondary bump can be seen in Fig.~\ref{fig:bump}. This
  notwithstanding, the participation ratio still vanishes
  asymptotically as
  \be Y_2(E) \sim \frac{1}{N}\,. \ee
 In analogy with recent results on   constraint-driven condensation~\cite{GB17}, we refer to this intermediate regime as the
  \emph{pseudo-condensate} phase.
\item[(C)] $E>E_c$: $\rho(\ve)$ has a bump placed at $\ve^* =
  \mycol{(E- E_{th})}$ and the participation ratio goes asymptotically
  to a value independent on $N$:
  \be
Y_2(E) =  \frac{(E-E_{\text{th}})^2}{E^2}\, .
  \ee
\end{itemize}
For finite and large $N$, it is thus possible to identify the
localization transition at $E_c$ by studying the scaling properties of
$Y_2(E)$ with $N$. We also observe that in the thermodynamic limit
$(E_c - E_{\text{th}})/N \rightarrow 0$ (see
Eq.~(\ref{eq:critical-E})), so that $Y_2(E/N)$ vanishes continuously
at $E_{\text{th}}/N$.

\subsection{Negative temperatures}
\label{sec:3-C}

\gcol{ Due to the results discussed above,} the only consistent
definition of temperature for energy values $E > E_{\text{th}}$ is the
microcanonical one,
\be
\frac{1}{T} = \frac{\partial S_N(A,E)}{\partial E}~.
\ee

\begin{widetext}
\gcol{ The behavior of $T(E)$ can be derived from the expression of
  the microcanonical entropy $S_N$ written in
  Eq.~(\ref{eq:rate-func-all}), taking also into account both the
  results on the matching regime and the known behaviours of
  $\mZ(\lambda,E)$ reported in Eqns.~(\ref{eq:ZE-Gaussian})
  and~(\ref{eq:ZE-asymptotic}). Explicitly, we obtain the following
  expression for $S_N(A,E)$ in the three regimes of interest:}
\begin{align}
  S_N(A,E)= \begin{cases} N s^{\text{back}} -
    (E-E_{\text{th}})^2/(2\sigma^2 N)~~\text{for}~~~~~~E-E_{\text{th}}
    \sim N^{1/2} \\ \\ N s^{\text{back}} - N^{1/3}\Psi(\zeta)
    ~~~~~~~~~~~~~~\text{for}~~~~~~E-E_{\text{th}} \sim N^{2/3} \\ \\ N
    s^{\text{back}} - \sqrt{E-E_{\text{th}}}
    ~~~~~~~~~~~~~~\text{for}~~~~~~E-E_{\text{th}} \sim N
\end{cases}.
\label{eq:entropies}
\end{align}
Accordingly, the microcanonical temperature reads:

\be  \frac{1}{T}= \begin{cases} -(E-E_{\text{th}})/(\sigma^2 N) ~~~~~~\text{for}~~~~~~E-E_{\text{th}} \sim N^{1/2}  \\ \\
 - N^{-1/3}  d\Psi(\zeta)/d\zeta ~~~~~~~~\text{for}~~~~~~E-E_{\text{th}} \sim N^{2/3} \\ \\
  - (2\sqrt{E-E_{\text{th}}})^{-1} ~~~~~~~~~\text{for}~~~~~~E-E_{\text{th}} \sim N
\end{cases}.
\label{eq:temp}
\ee
\end{widetext}
As a first important result, we find that the microcanonical
temperature is always negative for $E> E_{\text{th}}$ (note that the
function $\Psi(\zeta)$ is increasing monotonically and positive
definite). In this region, $T$ is `large' in absolute value, with a
scaling with the system size that depends on the energy scale. More
precisely, from Eq.~(\ref{eq:temp}) it follows that $T \sim N^{1/2}$
for $E-E_{\text{th}} \sim N^{1/2}$, $T \sim N^{1/3}$ for
$E-E_{\text{th}} \sim N^{2/3}$ and $T \sim N^{1/2}$ for
$E-E_{\text{th}} \sim N$.  This observation allows us to extend the
random phases approximation used to neglect hopping terms in the
Hamiltonian of the DNLSE to the whole region of the phase diagram in
Fig.~\ref{fig1} above the line $e_{th}=2a^2$.

Another important outcome follows from the comparison with the
localization properties discussed in Sec.~\ref{sec:3-B} and concerns
the possibility to observe {\it delocalized states with negative
  temperature} for $E_{\text{th}}<E<E_c$, i.e. in the
pseudo-condensate region~\cite{LETT}.


Such peculiar states are
nevertheless a finite-size effect. \gcol{In fact, 
the specific energy \emph{gap} where such states are equilibrium ones
shrinks as}
\be
e_c - e_{th} \sim {N^{-1/3}} \,,
\label{eq:gap}
\ee
where $e_c=E_c/N$. Still, a finite-size correction of the order
$N^{-1/3}$ is a relevant one, and this makes the pseudo-condensate
state accessible in realistic setups, where $N$ is not too large, as
it is typical in atomic condensates trapped in optical
lattices~\cite{TS,FLOP}. For example, the black dotted curve in
Fig.~\ref{fig1} represents the location of the critical energy density
$e_c(a)$ for a system with $N=100$ lattice sites, as obtained from
Eqs.~(\ref{eq:spinodal-zl}) and~(\ref{eq:crit-spin-ratio}).  The
associated pseudo-condensate region is thereby confined between this
line and the infinite-temperature line.

\begin{figure}[ht]
  \includegraphics[width=0.95\columnwidth]{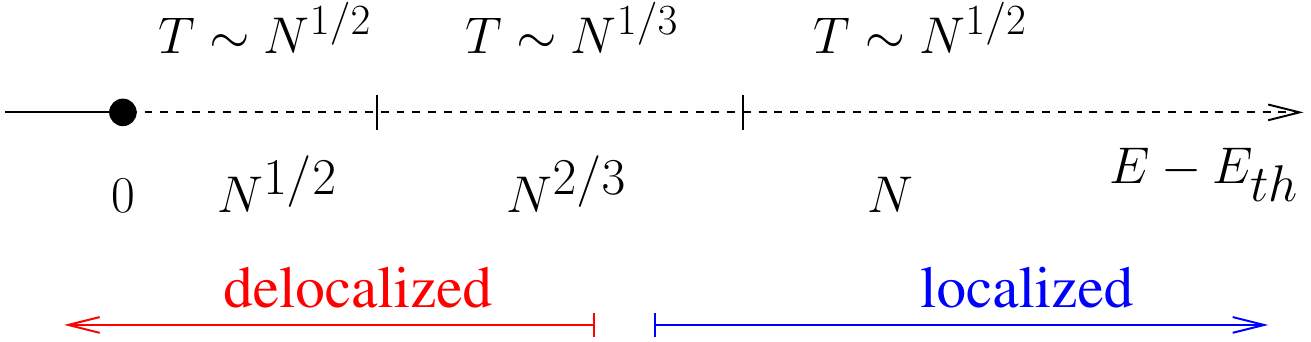}
  \caption{Finite-size scaling of the microcanonical temperature $T$
    for the different regimes of $E-E_{th}$, see Eq.~(\ref{eq:temp}).
    $T$ is negative in the whole semiaxis $E-E_{th}>0$, see dashed
    line.  The location of the localization transition is also shown
    qualitatively.}
\label{fig:temp}
\end{figure}

Altogether, the main results of this analysis are summarized in
Fig.~\ref{fig:temp}.

%

Let us finally point out that negative temperatures are not a
peculiarity of the non-equivalence between canonical and
microcanonical ensembles. They exist also in situations where the two
ensembles are equivalent, provided the Hamiltonian of the system is a
bounded function~\cite{CPV15,BPV18,MBV19}. On the other hand, in the
case of unbounded Hamiltonians, as in the case of the DNLSE, negative
temperatures have a physical meaning only within the microcanonical
ensemble, thus making ensemble inequivalence a necessary condition for
the observation of negative temperature states.

\subsection{Numerical simulations: the pseudo-condensate phase}
\label{sec:3-D}

In this subsection we focus on the equilibrium properties of the
pseudo condensate phase as obtained from numerical simulations.
Without any loss of generality, we have fixed $a=1$.  Accordingly,
from Eq.~(\ref{eq:moments-1-2}) one obtains $\langle \varepsilon
\rangle = 2$ and $\sigma^2=20$, respectively. This, in turn, yields
the following values for the spinodal point of the condensate and for
the transition point in the scaling variable $\zeta$, as shown in
Fig.~\ref{fig:transition}:
\bea
\zeta_l &\approx& 8.77 \nonumber \\
\zeta_c &\approx& 11.05
\eea
To well appreciate the pseudo-condensate region, we have focused on a
relatively small size, namely $N=128$. With this choice, the threshold
energy $\mycol{e}_{\text{th}}$ for ensemble inequivalence and the
localization energy $\mycol{e}_c$ reads off, accordingly, as:
\bea
\mycol{e}_{\text{th}} &=& 2 a^2 = 2 \nonumber \\
\mycol{e}_c &=& \mycol{e}_{\text{th}} + \frac{\zeta_c}{N^{1/3}} \approx 4.2
\label{eq:fsize-zc}
\eea
\begin{figure}[ht]
  \includegraphics[width=0.9\columnwidth]{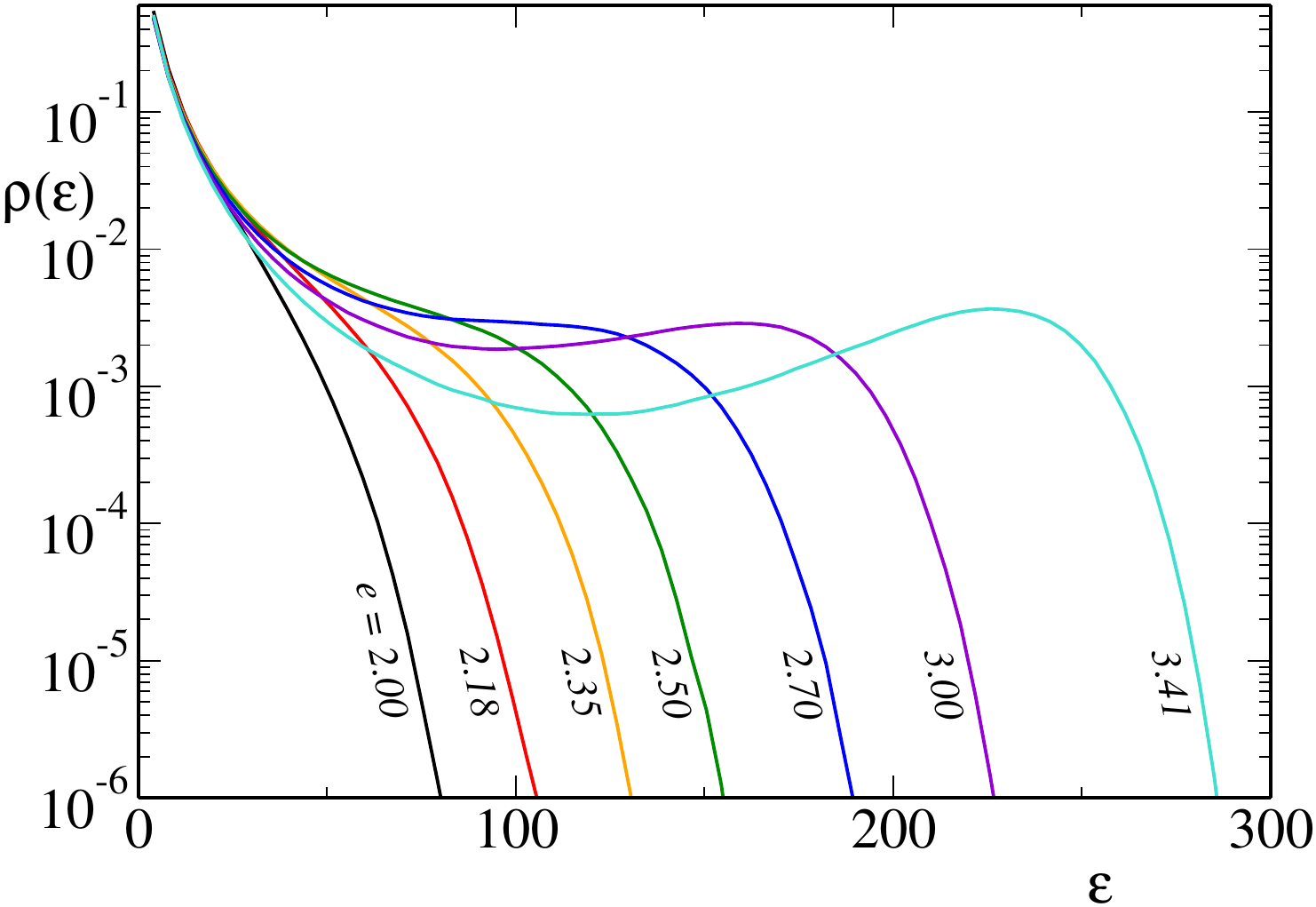}
  \caption{Marginal energy density probability $\rho(\ve)$ (see
      Eq.~(\ref{eq:marginal-ene})) obtained from numerical simulations
    for $N=128$, $a=1$ and different values of the energy density,
    namely $\mycol{e}=2.00 , 2.18, 2.35, 2.50, 2.70, 3.00,3.41 $. The
    threshold energy (per d.o.f.) for the data in figure is
    $\mycol{e_{\text{th}}}=2$ while the critical energy is
    $\mycol{e_c}=4.2$.}
\label{fig:bump}
\end{figure}

Microcanonical equilibrium simulations have been performed by evolving the system with a stochastic 
algorithm that was introduced in~\cite{IPP14} and used also
in~\cite{SEM14,SEM14b,IPP17} for the investigation of constraint-driven
condensation. 
This  algorithm allows to sample the phase space characterized by fixed values of the quantities: 

\be
A = \sum_{j=1}^N \rho_j^2 \qquad\qquad\qquad E = \sum_{j=1}^N \rho_j^4.
\ee
In detail, we have considered local random updates of randomly selected triplets of neighbouring
lattice sites $(j-1,j,j+1)$  such that the following constraints are satisfied:
\begin{widetext}
%
\bea
\rho_{j-1}^2(t+1) +   \rho_j^2(t+1)  +  \rho_{j+1}^2(t+1) ~~&=&~~ \rho_{j-1}^2(t) +   \rho_j^2(t)  +  \rho_{j+1}^2(t) \nonumber \\
\rho_{j-1}^4(t+1) +   \rho_j^4(t+1)  +  \rho_{j+1}^4(t+1) ~~&=&~~ \rho_{j-1}^4(t) +   \rho_j^4(t)  +  \rho_{j+1}^4(t), \nonumber \\
\eea
\end{widetext}
where the variable $t$ is the discrete time of the algorithm and is
measured in numbers of random moves divided by $N$.
To sample the equilibrium microcanonical measure of the model, 
we have evolved the system for a period of $2^{16}$ time units starting
from a partially localized initial condition~\footnote{For a fixed value of the 
total energy $E$ and system size $N=128$, we have chosen a set of initial conditions where  a fraction
$n_l=3/4 N$ of sites has a  value of local mass $\rho_j^2=a_l$ while in the remaining part 
of the system the local mass is $\rho_j^2=a_h$, with $a_h>a_l$. The parameters $a_l$ and $a_h$
are solutions of the equations $3 a_l + a_h=4$ and $3 a_l^2 + a_h^2=4E/N$. 
   }.
 By monitoring the evolution of the 
participation ratio $Y_2(E,t)$, we have verified that the relaxation transient was
sufficient to ensure relaxation to equilibrium in the range of energies $E$ here considered.
After this transient, we have  sampled the marginal distribution $\rho(\ve)$ for times up to
$t=2^{27}$ units.


Fig.~\ref{fig:bump} shows $\rho(\ve)$ for increasing values of the
total energy density $e<e_c$.  For sufficiently large values of $e$, a
bump is clearly visible in the region of large $\ve$.  Therefore,
despite the fact that the system is below the localization threshold, we observe
that the pseudo-condensate phase is characterized by a process of
partial energy localization. In other words, the non-monotonic decay
of $\rho(\ve)$ is not enough to conclude that the system is in the
localized phase. For this purpose, a finite-size study of $\rho(\ve)$
is necessary.

\section{Anomalous large-deviation exponent: physical meaning}

After the detailed exposition of all our results in the previous
sections, it is now worth to rationalize the physical meaning of the
anomalous large-deviation exponents we have found, i.e.  the exponent
$2/3$ for the matching regime and the exponent $1/2$ for the extreme
large-deviation regime. Where do these exponents come from? The answer
makes reference to some formulae already met in our calculations.  For
the sake of clarity, let us start with the exponent $1/2$.  The
behaviour of the entropy (in excess of the backgroud entropy
$S_N^{\text{back}}(a)$), i.e., $S_N-S_N^{\text{back}}(a) \sim -
N^{1/2} \sqrt{e-e_{th}}$ [see the third line of
  Eq. (\ref{eq:entropies})], for deviations of order $e-e_{th} \sim
1$, is the true landmark of localization. The exponent $1/2$ comes
directly from the interplay between the order of the non-linearity and
the conservation constraint on the norm of the \emph{wavefunction}.
In fact, in the microcanonical partition function one has to account
both for the conservation constraints on the sum of each variable
squared, i.e. $A=\sum_{j=1}^N |z_j|^2$, and on the energy
\be
E=\sum_{j=1}^N |z_j|^m
\label{eq:E-generic}
\ee
with $m>2$. As a consequence one obtains an expression of the partition
function in the form of  Eq.~(\ref{eq:semipart}),
where, for $m$ generic, the asymptotic decay of the individual probability is given by
\be
f_\lambda(\ve_j) \sim \exp(-\varepsilon^{2/m}_j).
\label{eq:individual-f-general}
\ee
As explained in Sec.~\ref{sec:2-A} and Sec.~\ref{sec:2-B} it is in
force of the asymptotic decay of $f_\lambda(\ve_j) $ in
Eq.~(\ref{eq:individual-f-general}) that the whole partition function
in the extreme large-deviation regime behaves (neglecting terms
independent of $E$) as:
\begin{align}
E - E_{th} \sim N &~~\Longrightarrow~~\Omega_N(A,E)\sim e^{- (E-E_{\text{th}})^{2/m}} \nonumber \\
& ~~\Longrightarrow~~\Omega_N(A,E)\sim e^{- N^{2/m} (e-e_{\text{th}})^{2/m}}
\end{align}
so that the total entropy, again neglecting terms which do not depend on $E$, reads
\begin{align}
& E - E_{th} \sim N ~~\Longrightarrow  \nonumber \\
& S_N(A,E)-S_N^{\text{back}}(a) \approx - N^{2/m} (e-e_{th})^{2/m} \nonumber \\
\end{align}
\gcol{Thus, for the DNLSE case, characterized by the value $m=4$, one
  has $S_N \sim N^{1/2}$.} While the generalization of the
large-deviation results reported in~\cite{MEZ05,SEM14,GM19} to the
case of generic $m$ is mathematically straightforward, we want to
emphasize its physical meaning: a detail of the microscopic
interactions of the system, i.e. the order of the non-linearity, is
revealed in the presence of localization at the macroscopic scale. It
is from the peculiar dependence of the entropy on $N$, or,
equivalently, from the dependence of the entropy on the total energy
that one can gather information on the microscopic interactions.\\

The way the exponent $2/3$ comes into play is very similar. It
also depends on the order of the non-linearity, though the connection
is less straightforward. On the one hand we can say that the value $1/3$ simply
comes from the matching between the Gaussian fluctuations and the
extreme large deviations of the partition function, as indicated in
Eq.~(\ref{eq:match}), which takes place at the scale $E-E_{th} \sim
N^{2/3}$, i.e. at $e-e_{th} \sim N^{-1/3}$
. Since the behaviour of extreme large deviations comes directly from
the order of the non-linear self-interaction~(\ref{eq:E-generic}),
while the shape of Gaussian fluctuations is independent from it, it is
easy to argue that the matching scale is solely dictated by the order
of the non-linearity. A longer pathway to reach the same conclusion
amounts to consider the expansion around $\beta=0$
($\beta\in\mathbb{C}$) of the ``local free-energy''
$\log[z(\lambda,\beta)]$ and to single out the scale where the
analytic and the non-analytic terms of the expansion are of the same
order~(see Sec.~\ref{sec:2-B}). As discussed in detail in
Sec.~\ref{sec:2-B}, such a scale turns out to be $\beta\sim N^{-1/3}$.

\section{Conclusions}
\label{sec:4}

In this paper we have shown how to compute the partition function of
the Non-Linear Schr\"odinger Hamiltonian [Eq.~(\ref{eq:Hamiltonian})]
in the microcanonical ensemble and close to the infinite temperature
line. This has allowed us to present a clear and coherent scenario for
the thermodynamics of this model, also in relation with its dynamical
properties. In fact, previous approaches (e.g., see \cite{RCKG00})
provided less transparent interpretations, due to the use of the
grand-canonical ensemble for describing also the phase above the line
$\beta = 0$ (see Fig.~\ref{fig1}). By making use of the microcanonical
approach, we have been able to show that this is a localized phase,
where typically a finite fraction of the whole energy is localized in
a few lattice sites. According to Ruelle~\cite{Ruelle}, this is
exactly one of the two conditions where equivalence between
statistical ensembles does not apply. Indeed, we have shown that the
$\beta =0$ line corresponds to the condition where the micro-canonical
and grand-canonical ensembles become inequivalent.

We also want to remark that, in a general perspective, 
the method we have used to compute the microcanonical partition function is
valid for lattices in \emph{any} spatial dimension.
There are further results emerging from our study, that merit to be
mentioned.

We have shown that for any finite value of the lattice sizes $N$ the
microcanonical temperature is negative for
$E>E_{\text{th}}$. Moreover, we have found that for finite $N$ the
localization transition is placed slightly above the $\beta =0$ line.
Consequently, there exists a region (the pseudo-condensate phase)
where one can observe negative-temperature states that are not
localized. This is a particularly important outcome in the perspective
of designing specific experiments of BEC in optical lattices, where
such peculiar states could be observed, while avoiding the
condensation of a large fraction of atoms onto a single or few sites
-- a condition that typically destabilizes the condensate.  Numerical
simulations performed in~\cite{CBSW} show that delocalized negative
temperature states can be observed also when a sufficiently high
thermal gradient is applied by two thermal reservoirs at positive
temperatures, coupled to the opposite ends of the DNLSE chain.  This
tells us that in out-of-equilibrium conditions, the standard thermal
phase and the localized one can coexist, thus confirming that the
microcanonical approach is consistent with the physics contained in
the DNLSE model.

Last, but not least, our analysis highlights how  finite-size
corrections to thermodynamic observables, like entropy and temperature,
are related, in the localized phase, to details of the microscopic
interactions.

Beside providing a deeper understanding of the thermodynamics of the
DNLSE, the results reported in this paper impact on wider
perspectives. \ggcol{We have highlighted the mechanism of localization
  in the DNLSE from a thermodynamic perspective, clarifying that it is
  driven exclusively by the interplay between global constraints and
  non-linear interactions.}

This result is significantly different form the typical phenomenon of
Many-Body Localization (MBL)~\cite{A18}, which is often framed in the
context of weakly perturbed integrable models whose quantum
unperturbed dynamics is characterized by the existence of as many
integrals of motion as degrees of freedom~\cite{RMS15}.

We point out that localization of the DNLSE is not due to disorder or
to the proximity in parameter space to a \ggcol{linear} model. The
high energy limit is in a sense an \emph{``effectively integrable
  limit''}, but the (independent) individual degrees of freedom are
non-linear oscillators.

This basic consideration at least challenges the generality of the MBL
approach, since localization in interacting-particle models can be
observed \ggcol{ in the present case without the need of imposing
  perturbatively small non-linearities.} Nonetheless, analogies with
the MBL approach are still traceable.  First of all, the DNLSE is a
genuine many-body model, because it represents a semiclassical limit
of the Bose-Hubbard Hamiltonian~\cite{FLOP}.  Second, in the
high-energy localized phase the DNLSE hamiltonian dynamics typically
drives the systems to a multi-breather state~\cite{FLOP,ICOPP,IFLOP}.
Each breather can be interpreted as the dynamical solution of an
integrable nonlinear oscillator of sufficiently high energy.  Even in
the presence of a background component, breathers can survive over
extremely long time lapses, due to their very weak interaction with
the lattice background and, accordingly, with any other breather.  On
the other hand, the number of breathers is not extensive and slowly
fluctuates in time~\cite{IFLOP}.

\section*{Acknowledgements}

We thank for interesting discussions on localization M. Baiesi,
S. Franz, L. Leuzzi, P. Marcati, G. Parisi, P. Politi,
F. Ricci-Tersenghi, L. Salasnich, A. Scardicchio, F. Seno,
A. Vulpiani. We also thank N. Smith for suggesting the argument for
$\zeta_c$ in Appendix~\ref{sec:app-D}. G.G. acknowledges the financial
support of the Simons Foundation (Grant No.~454949, Giorgio Parisi)
during the first period of this work. S.I. acknowledges support from
Progetto di Ricerca Dipartimentale BIRD173122/17 of the University of
Padova. R.L acknowledges partial support from project MIUR-PRIN2017
\emph{Coarse-grained description for non- equilibrium systems and
  transport phenomena} (CO-NEST) n. 201798CZL.

\section{Appendices}
\label{sec:app}

\subsection{Derivation of the rate function $\chi(\zeta)$ in the intermediate
matching regime}
\label{sec:app-A}

In this Appendix we illustrate how to obtain an estimate of the
integral  in Eq.~(\ref{eq:instanton}), making use of the saddle point
approximation, valid in the limit of large values of $N$.
We are dealing with the following expression
\be
\mycol{\mC(\lambda,\zeta)}= N \int_{\Gamma_{(+)}} \frac{d\beta}{i}\frac{1}{\sqrt{2\pi \langle \ve \rangle N^{1/3}\beta}}~e^{N^{1/3}F_\zeta(\lambda,\beta)}
\label{eq:anomalous-int-app}
\ee
where the non-negative quantity $\zeta$ is defined in Eq.(\ref{eq:scaling-variable}), and $\lambda$
can be assumed, for the sake of simplicity at this stage of the calculations, to be a positive free parameter.
In practice, we have to find under which conditions the function
\be
F_\zeta(\lambda,\beta) = \beta \zeta + \frac{10}{\lambda^4}\,\beta^2 +\frac{\lambda^2}{4\beta} \,.
\label{Fzs.1}
\ee
has a maximum for negative values of $\beta$, since the contour $\Gamma_{(+)}$, along which this
integral has to be evaluated, lies in the upper left quadrant of the complex $\beta$-plane.
At variance with Eq.(\ref{eq:Fz}), in the above expression of $F_\zeta(\lambda,\beta)$ we have made explicit
its dependence on $\lambda$, making use of Eqs. (\ref{eq:moments-1-2}).
For the sake of simplicity, let us introduce the shorthand notations $F_\zeta'(\lambda,\beta)$
and $F_\zeta''(\lambda,\beta)$ for the first and second derivatives of $F_\zeta(\lambda,\beta)$
with respect to $\beta$. The saddle point equation to be solved is
\begin{equation}
F_\zeta'(\lambda,\beta) = \zeta+ \frac{20}{\lambda^4}\,\beta -\frac{\lambda^2}{4\beta^2} = 0
\label{dfzs.1}
\end{equation}
It turns out that the position of the three roots of this equation in the complex $\beta$-plane
depend on the value of $\zeta$. More precisely, there  exists a specific value of this non-negative
parameter
 \begin{equation}
\zeta_l= 3 \cdot \frac{5^{2/3}}{\lambda^2}
\label{zl_exact.1}
\end{equation}
such that for $\zeta < \zeta_l$ there are  two
complex conjugate roots with negative real part and one real root on the positive axis, i.e. the saddle point equation does not provide
us with any acceptable solution. Conversely, for  $\zeta > \zeta_l$ there are three real roots $\beta_1 < \beta_2 < \beta_3$: the first two roots
are negative, while the third one is positive (therefore, uninteresting for our calculation).
\begin{figure}
  \includegraphics[width=0.9\columnwidth]{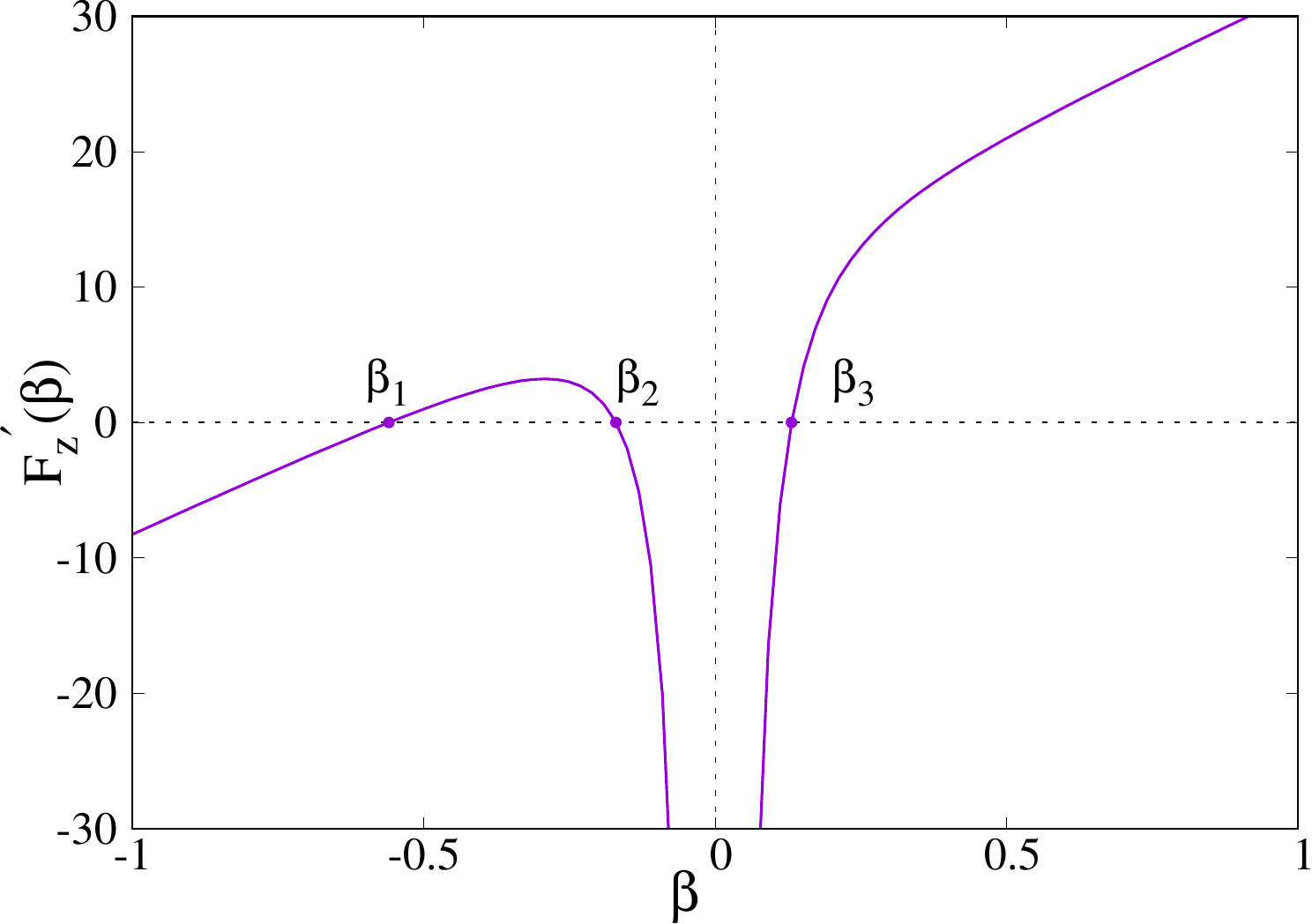}
\caption{Plot of $F_\zeta'(\lambda, \beta)$
 as a function of $\beta\in \mathbb{R}$
   for  $\lambda=1$ and $\zeta=12$.}
\label{fig:der}
\end{figure}

In order to make clear the latter scenario,
in Fig.~\ref{fig:der} we plot $F_\zeta'(\lambda,\beta)$ as a function of $\beta \in \mathbb{R}$ for $\lambda = 1$ and $\zeta = 12 > \zeta_l$.
Making reference to this picture we can explain the origin of $\zeta_l$: it identifies the bifurcation point where, by decreasing $\zeta$, the two negative
real roots $\beta_1$ and $\beta_2$ eventually coincide and then they become complex conjugate. In practice, the value of $\zeta_l$ can be obtained by
imposing the condition that the value of $F_\zeta'(\lambda,\beta)$ at its maximum, located in the negative value $\beta_m\in (\beta_1, \beta_2)$
vanishes. The value of $\beta_m = -\frac{\lambda^2}{2 \cdot 5^{1/3}}$ is obtained by imposing the condition $F_\zeta''(\lambda,\beta_m)
=0$, while $\zeta_l$ is finally determined by the condition $F_\zeta'(\lambda,\beta_m)=0$.
Another important outcome that we can obtain by direct visual inspection of Fig.\ref{fig:der} is that $F_z''(\lambda, \beta_1)>0$
and $F_z''(\lambda, \beta_2)<0$, from which we can conclude that  the maximum of $F_\zeta(\lambda,\beta)$ is at $\beta_2$,
which is the saddle-point solution that we are looking for.

Summarizing, we finally obtain the saddle-point estimate
\begin{equation}
\mycol{\mC(\lambda,\zeta)} \approx \exp[- N^{1/3} \chi(\zeta)]\
\label{saddle_s1.1}
\end{equation}
where the rate function $\chi(\zeta)$ is given by
\begin{equation}
\chi(\zeta) = - F_\zeta(\lambda, \beta=\beta_2)=-\beta_2\,\zeta-\frac{10}{\lambda^4}\beta_2^2- \frac{\lambda^2}{4\beta_2}
\label{chiz.B1}
\end{equation}
Making use of the saddle point equation~(\ref{dfzs.1}),  one can further
simplify the above equation as follows:
\begin{equation}
\chi(\zeta)= - \frac{\zeta\beta_2}{2}-\frac{3 \lambda^2}{8\beta_2}\, .
\label{chiz.2}
\end{equation}

\subsection{Properties of $\chi(\zeta)$}
\label{sec:app-B}

Here we describe the rate function
$\chi(\zeta)$ in the range $\zeta_l<\zeta<\infty$, where $\zeta_l$ is given in Eq.~(\ref{zl_exact.1}).  As discussed in the previous Appendix, for $\zeta =\zeta_l$
the roots $\beta_1$ and $\beta_2$ coincide with $\beta_m = -\frac{\lambda^2}{2 \cdot 5^{1/3}}$.
Substituting this value  into Eq.~(\ref{chiz.2}), one obtains
\begin{equation}
\chi(\zeta_l) = \frac{3}{2}\, \left(\frac{10}{\lambda^2}\right)^{2/3}\,,
\label{chiz_ll}
\end{equation}
as reported in the first line of
Eq.~(\ref{eq:chi-asymptotics-intro}).

Finding the  behavior of $\chi(\zeta)$ for large values of $\zeta$ can be easily obtained making
use of a suitable reparametrization of the root $\beta_2$ in the form
\be
\beta_2= -\frac{\lambda}{2\sqrt{\zeta}} \theta_\zeta \,.
\label{eq:rescale-sz}
\ee
Then Eq.~(\ref{dfzs.1}) can be rewritten in the form
\be
-b(\zeta)~\theta_\zeta^3+\theta_\zeta^2-1 =0,
\label{eq:sad-eq-pert}
\ee
where
\be
b(\zeta) = \frac{10}{\lambda^3} \frac{1}{\zeta^{3/2}}.
\label{eq:bz-def}
\ee
Note that, due to the change of sign in passing from $\beta_2$ to $\theta_\zeta$, we have
to find the smallest positive root $\theta_\zeta$ in Eq.~(\ref{eq:sad-eq-pert}). Also Eq.~(\ref{chiz.2})
can be rewritten in terms of $\theta_\zeta$ as follows
\begin{equation}
\chi(\zeta)= \sqrt{\zeta}~\frac{\lambda}{4}~\frac{\theta_\zeta^2+3}{\theta_\zeta}\, .
\label{eq:chi-useful}
\end{equation} \\

From Eqs.~(\ref{eq:sad-eq-pert}) and (\ref{eq:bz-def}) it
follows immediately that $\theta_\zeta\to 1$ for $\zeta\to \infty$. Hence, for
large $\zeta$ (or equivalently small $b(\zeta)$), we obtain a
perturbative solution of Eq.~(\ref{eq:sad-eq-pert}), which at leading order reads
\begin{equation}
\theta_\zeta= 1+ \frac{b(\zeta)}{2} + \mathcal{O}\left({b(\zeta)}^2\right)\, .
\label{pert_thetaz.1}
\end{equation}
By substituting this perturbative solution
into Eq.~(\ref{eq:chi-useful}) we finally obtain the
behavior of $\chi(\zeta)$ for large values of $\zeta$ in the form
\be
\chi(\zeta) = \lambda~\sqrt{\zeta}-\frac{5}{2 \lambda^2}\frac{1}{\zeta} + \mO\left(\frac{1}{\zeta^{5/2}}\right)\,,
\label{eq:rate-function-with-theta}
\ee
as reported in the second line of Eq.~(\ref{eq:chi-asymptotics-intro}).

More generally, the explicit expression of the smallest positive root of Eq.~(\ref{eq:sad-eq-pert}) can be obtained making use
of {\it Mathematica}; it reads:
 \begin{widetext}
\begin{equation}
\theta_\zeta =  \frac{1}{3b_\zeta} + \frac{1}{3\cdot 2^{2/3} b_\zeta} \frac{(1-i\sqrt{3})}{\left( -2
      + 27 b_\zeta^2 + 3 \sqrt{-12+81 b_\zeta^2} \right)^{1/3}}  +
\frac{1}{3\cdot 2^{4/3} b_\zeta}(1+i\sqrt{3}) \left( -2 + 27 b_\zeta^2 + 3 \sqrt{-12+81 b_\zeta^2} \right)^{1/3}
\label{thetaz_math1}
\end{equation}
\end{widetext}
where we have introduced the shorthand notation $b_\zeta$ for $b(\zeta)$.
By making use of Eq.~(\ref{zl_exact.1}), we can conveniently rewrite  $b_\zeta$ in the
dimensionless form
\begin{equation}
b_\zeta^2 = \frac{1}{2}~\left( \frac{2}{3 \, r}\right)^3 \, ,
\label{bz2}
\end{equation}
where  $r=\zeta/\zeta_l\ge 1$.
Accordingly, we can also rewrite the saddle-point solution ~(\ref{thetaz_math1}) in the form
\begin{equation}
\theta_\zeta \equiv \theta(r)= \frac{\sqrt{3}}{4} r^{3/2} \left[ 2 + \frac{(1-i\sqrt{3})}{g(r)} +
  (1+i\sqrt{3})g(r) \right]
\end{equation}
where
\begin{equation}
g(r) = \frac{1}{r} \left( 1 + i ~\sqrt{r^3-1} \right)^{2/3}.
\end{equation}
By multiplying both the numerator and the denominator of $\theta(r)$ by $(1-i
~\sqrt{r^3-1})^{2/3}$
one obtains the following expression for the saddle-point solution:
\begin{equation}
\theta(r) = \frac{\sqrt{3}}{4} r^{3/2} \left[ 2 + \frac{1}{r}\left(
  \xi~u_r^{2/3}+\overline{\xi}~\overline{u}_r^{2/3} \right)  \right],
\label{eq:thetar-complex}
\end{equation}
where $\xi$ and $u_r$ denote, respectively, a complex number and a
complex function of the real variable $r$:
\begin{eqnarray}
\xi &=& 1 + i \sqrt{3} \nonumber \\
u_r &=& 1 + i ~\sqrt{r^3-1}.
\label{eq:complex-exp}
\end{eqnarray}
The overbars in Eq.(\ref{eq:thetar-complex}) denote the complex conjugate quantities.
After some straightforward algebra, one obtains the following expression of
the saddle-point solution
\begin{equation}
  \theta(r) = \frac{\sqrt{3}}{2} r^{3/2} \left[ 1 + 2 \cos\left( \frac{\pi}{3} + \frac{2}{3} \textrm{arctan}(\sqrt{r^3-1})\right) \right]
\label{eq:theta-r}
\end{equation}
Summarizing all these calculations, we finally observe that from Eq.(\ref{eq:chi-useful}) it follows that the rate function
$\chi(\zeta)$ can be written explicitly in terms of this explicit saddle-point solution.

\subsection{The critical value $\zeta_c$}
\label{sec:app-D}

In this appendix we obtain the critical value $\zeta_c$, which
is identified by the \emph{matching condition}  $\chi(\zeta_c) = \zeta_c^2/(2\sigma^2)$, i.e., the value at which
the two branches in Fig.~\ref{fig:transition} cross each other.
Making use of the definitions introduced in the previous Appendix we can write explicitly
the matching condition in the  form
\begin{equation}
\sqrt{\zeta_c} \frac{\lambda}{4}\,\left[ \frac{\theta(r_c)^2 +3}{\theta(r_c)}\right]= \frac{\lambda^4}{40}\zeta_c^2\, ,
\label{zc.1}
\end{equation}
where $r_c = \zeta_c/\zeta_l$. Taking into account Eq.(\ref{zl_exact.1}), the previous equation
can be rewritten as follows:
\begin{equation}
\frac{\theta^2(r_c)+3}{\theta(r_c)}= \frac{3^{3/2}}{2}\, r_c^{3/2}\, .
\label{rc.1}
\end{equation}
A shortcut to obtain the expected result amounts to substitute this relation directly into the
equation
\begin{equation}
-\frac{1}{\sqrt{2}}\, \left(\frac{2}{3}\right)^{3/2}\, r^{-3/2}\, \theta(r)^3 + \theta(r)^2 -1=0\, ,
\label{theta_r}
\end{equation}
which is a suitable way of rewriting the saddle-point equation (\ref{eq:sad-eq-pert}) in terms of  the adimensional variable $r$.
Thus one obtains the simple result
\begin{equation}
\theta(r_c)= \sqrt{\frac{3}{2}}\, .
\label{thetarc}
\end{equation}
As a final step it remains to substitute this result into Eq.(\ref{rc.1}) and we finally obtain
\begin{equation}
r_c= \frac{\zeta_c}{\zeta_l}= 2^{1/3}
\label{exact_rc}
\end{equation}
One can obtain an immediate check of this result by computing from Eq.~(\ref{eq:theta-r})
$\theta(r_c) = \sqrt{\frac{3}{2}}$, as expected. \\

\vskip 0.4cm
\bibliography{biblio}

\end{document}